\documentclass[pre,twocolumn,aps,groupedaddress]{revtex4}
\usepackage[T1]{fontenc}
\usepackage[latin9]{inputenc}
\usepackage{array}
\usepackage{booktabs}
\usepackage{mathrsfs}
\usepackage{multirow}
\usepackage{amsmath}
\usepackage{amsthm}
\usepackage{amssymb}
\usepackage{mathbbol}
\usepackage{stmaryrd}
\usepackage{graphicx}
\usepackage{latexsym}
\usepackage{textcomp}
\usepackage{mathtools}
\usepackage{xcolor}
\usepackage[citecolor=magenta,colorlinks=true]{hyperref}
\usepackage{lineno}
\usepackage{stackrel}
\usepackage[toc,page,titletoc]{appendix}
\usepackage{bm}
\definecolor{mycolor1}{rgb}{0.1, 0.6, 0.6}
\newcommand{\beginsupplement}{%
        \setcounter{table}{0}
        \renewcommand{\thetable}{S\arabic{table}}%
        \setcounter{figure}{0}
        \renewcommand{\thefigure}{S\arabic{figure}}%
         \setcounter{equation}{0}
        \renewcommand{\theequation}{S\arabic{equation}}%
     }
\begin{document}
\title{Long-range correlations in pinned athermal networks}

\author{Debankur Das}
\email{debankurd@tifrh.res.in}
\affiliation{Centre for Interdisciplinary Sciences, Tata Institute of Fundamental Research, Hyderabad 500107, India}
\author{Pappu Acharya}
\email{pappuacharya@tifrh.res.in}
\affiliation{Centre for Interdisciplinary Sciences, Tata Institute of Fundamental Research, Hyderabad 500107, India}
\author{Kabir Ramola}
\email{kramola@tifrh.res.in}
\affiliation{Centre for Interdisciplinary Sciences, Tata Institute of Fundamental Research, Hyderabad 500107, India}
\date{\today}

\begin{abstract}
We derive exact results for displacement fields that develop as a response to external pinning forces in two dimensional athermal networks. For a triangular lattice arrangement of particles interacting through soft potentials, we develop a Green's function formalism which we use to derive exact results for displacement fields produced by localized external forces. We show that in the continuum limit the displacement fields decay as $1/r$ at large distances $r$ away from a force dipole. Finally, we extend our formulation to study correlations in the displacement fields produced by the external pinning forces. We show that uncorrelated pinned forces at each vertex give rise to long-range correlations in displacements in athermal systems, with a non-trivial system size dependence. We verify our predictions with numerical simulations of athermal networks in two dimensions.
\end{abstract}
\maketitle

 \section{Introduction} 
 
 Networks composed of athermal constituents such as jammed particles arise in a variety of contexts \cite{o2003jamming}, including in granular \cite{behringer2018physics,goldenberg2002force} and glassy systems \cite{Ikeda}, active matter \cite{henkes}, as well as biological tissues \cite{dapeng1,broedersz2011criticality,boromand2018jamming}. Such materials are robust to thermal agitations and differ from thermal systems in their response to external perturbations, as well as fluctuations in positions and forces \cite{acharya}. 
The response of athermal materials to external perturbations has many industrial as well as biological applications \cite{cates1998jamming,head2005mechanical,athanassiadis2014particle,Ramola,philippe,geng}, and continues to be the subject of active research. 
Similarly, athermal systems driven by {\it local} active forces arise in various contexts in physics and biology \cite{janevs2019statistical,janevs2019statistical1,ronceray2016fiber,ronceray2019fiber,schwarz2013physics}. 
Although many properties of athermal systems have been extensively studied over the last few decades, the fluctuations in displacement fields, as well as the long-range correlations that develop in such systems, are relatively less well understood \cite{dapeng2,degiuli2018field,lemaitre2014structural,lemaitre2021stress,chikkadi2011long}. Developing theories for the collective behaviour of athermal systems in the presence of external forces such as gravity or active internal forces therefore represents a new challenge \cite{kadanoff1999built,de1999granular}. 

The quasi-static response of athermal materials to local pinning forces is also important in the study of granular materials as well as glasses \cite{bhowmik2019effect,zeng1999absence,lerner2018characteristic}, where local force perturbations can be used to extract lengthscales \cite{rainone2020pinching}. In such systems, the constraints of mechanical equilibrium alone do not provide enough equations to solve for the stress tensor, which can lead to non-trivial stress transmission properties \cite{liu1995force,jishnu}. Although theories of continuum elasticity posit constitutive relations between the microscopic stress and strain fields, it is as yet unclear how such relationships emerge at large lengthscales in disordered athermal materials. For example, the continuum equations that emerge can be elliptic, or hyperbolic \cite{bouchaud2002granular}, depending on the nature of the underlying medium, but a clear understanding of this phenomenon is still lacking. Moreover, predicting fluctuations and correlations in athermal ensembles remains a challenging theoretical task. In this context it is useful to appeal to systems where the strain field can be computed exactly to determine the nature of the correlations and response in athermal systems .

\begin{figure}[t!]
\includegraphics[scale=0.4]{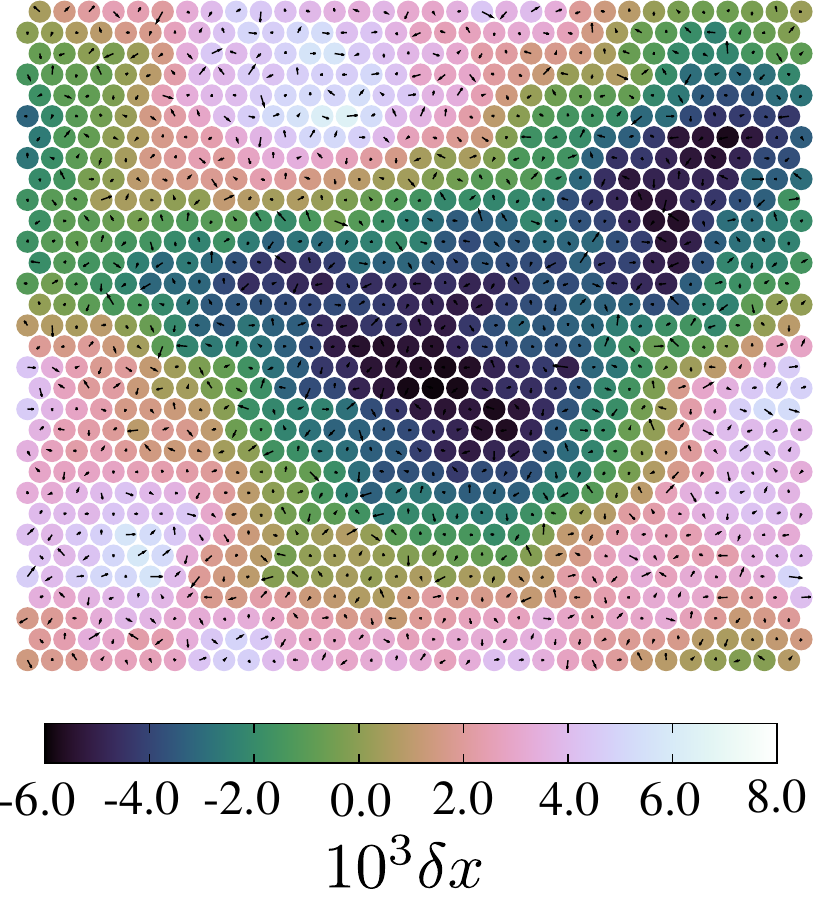}
\caption{The displacement response of an athermal membrane composed of soft particles to the presence of active (external) forces imposed on each site (depicted with arrows). The particles are colored according to their displacements from the crystalline positions along the $x$-direction. The external forces are drawn from uncorrelated underlying distributions. The displacements of the sites from their crystalline positions as a response, displaying large correlated regions. The separation between the particles in the initial crystalline state is $R_0 = 1.2$, and the equilibrium bond lengths are $L_{\text{rest}} = 1.1$.
}
\label{fig_schematic}
\end{figure}

The stress response of triangular networks with microscopic force balance constraints has been a paradigmatic model of stress transmission in granular systems \cite{liu1995force,Snoeijer,farhang}, and has been used to model continuum athermal elasticity at large lengthscales \cite{otto2003anisotropy}. However, incorporating the effects of microscopic disorder, such as in the external forces, within an exact framework has remained difficult owing to the non-trivial spatial arrangements of particles in minimum energy configurations. In this context it is important to appeal to systems where exact results can be obtained. 
In this paper we present exact results for displacement fields and their correlations in athermal networks using a model of frictionless soft particles in an initial triangular lattice arrangement. 
Our exact results demonstrate that in the athermal systems where mechanical equilibrium is {\it exactly} imposed at the local level, uncorrelated external forces can give rise to large correlated regions in the system. Indeed, as we show, in addition to being long-ranged, the displacement correlations have a non-trivial system size dependence.

\section{Pinned Network Model}
\label{section_model}

We study a system of equal-sized particles with initial positions $\{\vec{r}_{i,0}\} \equiv \{x_{i,0},y_{i,0}\}$, arranged in an $L \times L$ triangular lattice with lattice constant $R_0$. We impose periodic boundary conditions in both the $x$ and $y$ directions. Each particle interacts with its nearest neighbours through a distance dependent force law. We consider these interactions to be harmonic, with a spring constant $K$ and an equilibrium bond length $L_{\text{rest}}$. Our results can be easily generalized to other types of interactions as well. The Hamiltonian of the system is given by
\begin{equation}
\mathcal{H} = \sum_{i=1}^{L^2} \frac{p_{i}^2}{2m_i} + \frac{K}{2} \sum_{i=1}^{L^2} \sum_{\langle i j \rangle} (|\vec{r}_i -\vec{r}_j| - L_{\text{rest}})^2,
\label{Eq_hamiltonian}
\end{equation}
where $m_i$ is the mass and $\vec{r}_{i}$ represents the instantaneous position of the $i^{\text{th}}$ particle. The brackets $\langle \rangle$ in the above summation denote nearest-neighbours on the triangular lattice network with $j>i$. We define $R_0 = L_{\text{rest}}(1 +  \alpha)$, where $\alpha$ quantifies the compression of the initial crystalline state. When $\alpha < 0$ the system is overcompressed and the forces between the vertices (particles) are repulsive, whereas when $\alpha > 0$ the system is under-compressed and the forces are attractive. We consider the athermal version of this system, i.e. the zero temperature limit in which the momentum term in Eq.~(\ref{Eq_hamiltonian}) is irrelevant, and we only deal with the {\it minimum} of the potential energy. In addition to the inter-particle forces, we impose forces $\vec{f}_{i,\text{ext}}$ at every vertex $i$ that represent the external pinning forces acting on the system. We display a typical force balanced configuration of such a network in the presence of random pinning forces in Fig.~\ref{fig_schematic}.

\section{Simulation Details}
\label{sec:Numerical}
In order to verify that the predictions from our theory are able to capture the non-trivial nature of the response in such systems, we simulate the athermal triangular network in the presence of external forces. We consider an ideal triangular lattice with lattice parameter $R_0 \neq L_{\text{rest}}$. At every vertex $i$, we impose an external force $\vec{f}_{i,\text{ext}}$. We consider force balanced configurations, i.e. configurations at energy minima. We minimize the  energy of the system using the FIRE (Fast Inertial Relaxation Engine) algorithm \cite{fire}, which can naturally incorporate externally imposed forces. The implementation of the algorithm is simple and rapidly leads to a minimum energy configuration. At every time step we compute the power $P = \vec{F}.\vec{v}$ in the entire system. If $P > 0$, the velocity is set to $\vec{v} \to (1 - \beta) \vec{v} + \beta \hat{F}|\vec{v}| $, the time step is increased as $\Delta t = \Delta t f_{\text{inc}}$, up to the maximum value $\Delta t = \Delta t_{\text{max}}$ and $\beta$ is changed to $\beta f_{\beta}$. If $P < 0$, the velocity is set to zero, the time step is decreased $\Delta t = \Delta t f_{\text{dec}}$ and $\beta$ is reset back to its initial value $\beta _{\text{start}}$. In our simulations, we use $\beta = \beta_{\text{start}} = 0.01$, $\Delta t = 0.0001$, $\Delta t_{\text{max}} = 0.001$, $f_{\beta} = 0.99$, $f_{\text{inc}} = 1.1$, and $f_{\text{dec}} = 0.5$.

\section{Linearized Force Balance}
\label{sec_linear_force_balance}

\begin{figure}[t!]
\includegraphics[width=0.6\linewidth]{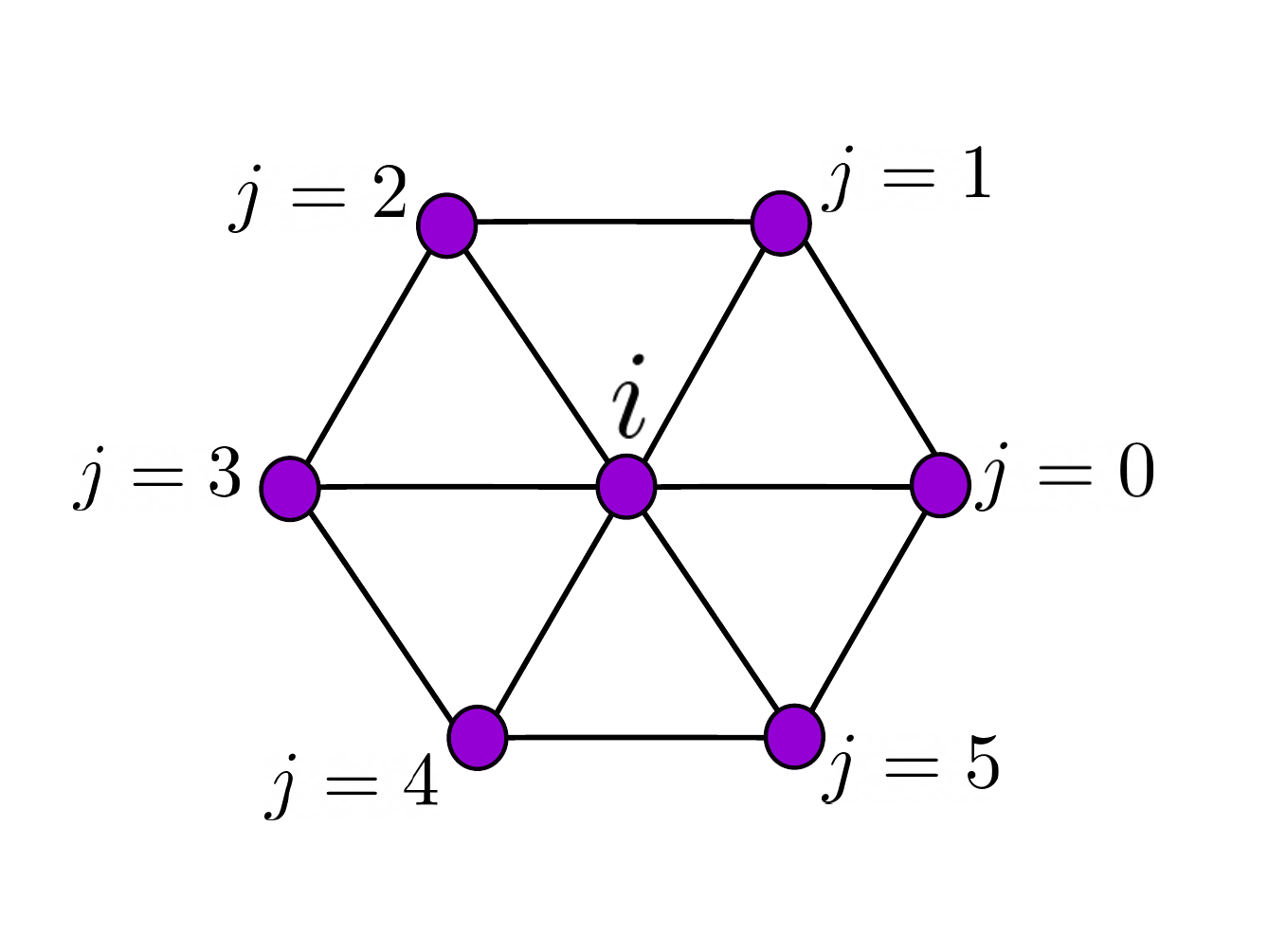}
\caption{The labeling convention used in our computation. The six neighbours of every node $i$ of the lattice are labeled $j = 0$ to $5$. The angles of the bonds between nodes in the reference crystalline state can take any of six values (depending on $j$) with the positive $x$-axis, $\theta_{ij}= 2 \pi j /6. $}
\label{Fig_lattice_figure}
\end{figure}

We begin by analyzing the response of the ideal triangular lattice in the limit of weak external forces. The inter-particle forces are determined from Eq.~(\ref{Eq_hamiltonian}) and are given by
\begin{equation}
\label{Eq_force_eq}
\begin{split}
f_{ij}^{x} = -K\left(\sqrt{x_{ij}^2 + y_{ij}^2} - L_{\text{rest}}\right)\frac{x_{ij}}{r_{ij}},
\\
f_{ij}^{y} = -K\left(\sqrt{x_{ij}^2 + y_{ij}^2} - L_{\text{rest}}\right)\frac{y_{ij}}{r_{ij}}.
\end{split}
\end{equation}
Here $f_{ij}^{x(y)}$ are the $x(y)$ components of the force between nodes $i$ and $j$ and $x_{ij} =  x_j -  x_i$ represents the distance between particles $i$ and $j$. The ground state of the system is determined by the condition of mechanical equilibrium, i.e. each site is in force balance with
\begin{equation}
\label{Eq_force_bal}
     \sum_{j=0}^{5} f_{ij}^{x} + {f}_{i,\text{ext}}^{x} = 0,~~~~~
     \sum_{j=0}^{5} f_{ij}^{y} + {f}_{i,\text{ext}}^{y} = 0, ~~~~~ \forall~ i,
\end{equation}
where the sum includes the six neighbours $j = 0$ to $5$ of the $i^{th}$ site (Fig. \ref{Fig_lattice_figure}). In addition to the above force balance equations at every site, we also impose {\it global} force balance on the system with 

\begin{equation}
\label{Eq_force_bal_global}
     \sum_{i = 1}^{L^2} {f}_{i,\text{ext}}^{x} = 0,~~~~~
     \sum_{i=1}^{L^2} {f}_{i,\text{ext}}^{y} = 0.
\end{equation}

In the absence of external forces, the ground state configuration is the unperturbed triangular lattice with particle positions $\{\vec{r}_{i,0}\}$. We next treat the introduction of external forces as a perturbation. As a response to this perturbation, the positions of the vertices change as
\begin{eqnarray}
\nonumber
\label{Eq_disp}
x_{i} &=& x_{i,0} + \delta x_{i},\\
y_{i} &=& y_{i,0} + \delta y_{i}.
\end{eqnarray}
Here $\delta x_{i}$ and $\delta y_{i}$ are the $x$ and $y$ displacements of the $i^{th}$ site from their positions in the initial triangular lattice. The force law in Eq.~(\ref{Eq_force_eq}) is a non-linear function of the inter-particle distances $x_{ij}$ and $y_{ij}$. In the limit of small perturbations, we can expand Eq.~(\ref{Eq_force_eq}) up to linear order in the relative displacements $\delta x(y)_{ij} = \delta x(y)_j - \delta x(y)_i $, leading to
\begin{equation}
\label{Eq_force_linear}
\begin{split}
    \delta f_{ij}^{x} = C_{ij}^{xx}\delta x_{ij} + C_{ij}^{xy}\delta y_{ij}, 
    \\
    \delta f_{ij}^{y} = C_{ij}^{yx}\delta x_{ij} + C_{ij}^{yy}\delta y_{ij}.
\end{split}
\end{equation}
Without loss of generality we may set $K=1$. The linear coefficients $C_{ij}^{\mu \nu}$ are then given by
   \begin{eqnarray}
   \nonumber
   C_{ij}^{xx} &=& \frac{-2R_0 + L_{\text{rest}} -L_{\text{rest}}\cos(\frac{2j\pi}{3})}{2R_0}, \\
   \nonumber
    C_{ij}^{xy} &=& -\frac{L_{\text{rest}}\sin(\frac{2j\pi}{3})}{2R_0}, \\
    \nonumber
   C_{ij}^{yx} &=& -\frac{L_{\text{rest}}\sin(\frac{2j\pi}{3})}{2R_0}, \\
   C_{ij}^{yy} &=& \frac{-2R_0 + L_{\text{rest}} + L_{\text{rest}}\cos(\frac{2j\pi}{3})}{2R_0}.
   \label{Eq_C-coeff}
   \end{eqnarray}
We note that the coefficients $C_{ij}^{\mu \nu}$ represent the elements of the Hessian matrix of the crystalline system. Crucially, these coefficients $C^{\mu \nu}_{ij}$, being drawn from the unperturbed crystalline arrangement are translationally invariant, i.e. they do not depend on the site index $i$. Next, using these linearized expressions we can relate the changes in positions to the external forces as
\begin{equation}
\begin{aligned}
\label{Eq_linear_force_bal}
    \sum_{j = 0}^5 C_{ij}^{xx}(\delta x_{j} - \delta x_{i}) + \sum_{j = 0}^5 C_{ij}^{xy}(\delta y_{j} - \delta y_{i}) =  -f_{i,\text{ext}}^x,  \\
    \sum_{j = 0}^5 C_{ij}^{yx}(\delta x_{j} - \delta x_{i}) + \sum_{j = 0}^5 C_{ij}^{yy}(\delta y_{j} - \delta y_{i}) =  -f_{i,\text{ext}}^y.  \\
    \end{aligned}
\end{equation}
These translationally invariant equations of force balance can be simplified in Fourier space. In order to define a Fourier transform we assign the displacement field $(\delta x_i, \delta y_i) \equiv (\delta x(\vec{r}), \delta y(\vec{r}))$ to every site $i$ at the {\it unperturbed} lattice positions $\vec{r} \equiv \vec{r}_{i,0}$. The Fourier transform of the displacements is $\delta \tilde{x}(\tilde{y}) (\vec{k}) = \sum_{\vec{r}} \exp(i\vec{k} . \vec{r}) \delta x(y)(\vec{r})$ and external forces is ${\Tilde{f}}^{x (y)}_{\text{ext}}(\vec{k}) =\sum_{\vec{r}}  \exp(i \vec{k} . \vec{r}) f_{i,\text{ext}}^{x(y)}$. Here $\vec{k} = (k_x,k_y) \equiv \Big(\frac{2\pi l}{2 L},\frac{2 \pi m}{L}\Big)$ are the reciprocal lattice vectors of the triangular lattice and the volume of the system is $V = 2 L^2$ \cite{horiguchi}. It is also convenient to define the basic translation coefficients in Fourier space (see Supplemental Material for details ~\cite{supplemental})
\begin{eqnarray}
\mathcal{F}_j(\vec{k}) &=& \exp(- i \vec{k}.\vec{\mathbb{r}}_j),
\end{eqnarray} 
where $\vec{\mathbb{r}}_j$ represent the lattice translation vectors given by $\vec{\mathbb{r}}_0 = (2,0)$, $\vec{\mathbb{r}}_1 = (1,1)$, $\vec{\mathbb{r}}_2 = (-1,1)$, $\vec{\mathbb{r}}_3 = (-2,0)$, $\vec{\mathbb{r}}_4 = (-1,-1)$, $\vec{\mathbb{r}}_5 = (1,-1)$. Next, multiplying Eq.~(\ref{Eq_linear_force_bal}) by $\exp(i\vec{k}.\vec{r})$ and summing over all sites of the lattice, we arrive at the following matrix equation at each reciprocal lattice point
\begin{equation}
\left(
\begin{matrix} 
A^{xx}(\vec{k}) & A^{xy}(\vec{k}) \\
A^{yx}(\vec{k}) & A^{yy}(\vec{k})
\end{matrix}
\right)
\left(
\begin{matrix} 
\delta \tilde{x} (\vec{k}) \\
\delta \tilde{y} (\vec{k})
\end{matrix}
\right)
= 
\left(
\begin{matrix} 
{-\tilde{f}}^{x}_{\text{ext}}(\vec{k}) \\
{-\tilde{f}}^{y}_{\text{ext}}(\vec{k})
\end{matrix}
\right).
\label{Eq_matrix_eq}
\end{equation}
The above matrix elements $A^{\mu \nu}$ can be expressed in terms of the coefficients $C_{ij}^{\mu \nu}$ as
\begin{eqnarray}
\label{Eq_A-matrix}
  \nonumber
     A^{xx}(\vec{k}) &=& -\sum_{j=0}^{5} (1- \mathcal{F}_j(\vec{k})) C^{xx}_{ij} ,\\
     \nonumber
     A^{xy}(\vec{k}) &=& -\sum_{j=0}^{5} (1 - \mathcal{F}_j(\vec{k})) C^{xy}_{ij},\\
     \nonumber
     A^{yx}(\vec{k}) &=& -\sum_{j=0}^{5} (1 - \mathcal{F}_j(\vec{k})) C^{yx}_{ij},\\
     A^{yy}(\vec{k}) &=&
     -\sum_{j=0}^{5} (1 - \mathcal{F}_j(\vec{k})) C^{yy}_{ij}.
\end{eqnarray}

The solution for the displacements in Fourier space in response to externally imposed forces can then be obtained by solving Eq.~(\ref{Eq_matrix_eq}). 

\section{Response Green's Functions}
\label{sec_Green_func}

We can interpret the matrix elements of $A^{-1}$ as Green's functions in Fourier space as
\begin{equation}
G = A^{-1} =
\left(
 \begin{matrix} 
\Tilde{G}_{xx}(\vec{k}) & \Tilde{G}_{xy}(\vec{k}) \\
\Tilde{G}_{yx}(\vec{k}) & \Tilde{G}_{yy}(\vec{k})
\end{matrix}
\right).
\label{Eq_inverse_matrix}
\end{equation}
These Green's functions can then be used to derive the Fourier transformed displacements as
\begin{eqnarray}
\nonumber
    \delta \tilde{x}(\vec{k}) &=& -\Tilde{G}_{xx}(\vec{k})\Tilde{f}^x_{\text{ext}}(\vec{k}) - \Tilde{G}_{xy}(\vec{k}) \Tilde{f}^y_{\text{ext}}(\vec{k}),\\
   \delta \tilde{y}(\vec{k}) &=& -\Tilde{G}_{yx}(\vec{k})\Tilde{f}^x_{\text{ext}}(\vec{k}) - \Tilde{G}_{yy}(\vec{k}) \Tilde{f}^y_{\text{ext}}(\vec{k}).
 \label{Eq_Fourier_solution}
\end{eqnarray}
The external forces therefore play the role of a source term that generate the displacement fields at every site. In order to obtain the actual displacements up to linear order, we perform an inverse Fourier transform of Eq.~(\ref{Eq_Fourier_solution}). Defining the Green's function in real space $\vec{r} = (x,y)$ as
\begin{equation}
\label{Eq_Green_func_real1}
G_{\mu \nu}(\vec{r}) =  \frac{1}{V} \sum_{l = 0}^{2L - 1} \sum_{m = 0}^{L - 1} e^{-i\vec{k}\cdot \vec{r}}  \tilde{G}_{\mu \nu}(\vec{k}),
\end{equation}

we arrive at the following form of the displacement field in real space
\begin{small}
\begin{equation}
 \begin{aligned}
\delta x(\vec{r}) = -\sum_{\vec{r}'} \left[ G_{xx}(\vec{r} - \vec{r}') \delta f^{x}_{\text{ext}}(\vec{r}') + G_{xy}(\vec{r} - \vec{r}') \delta f^{y}_{\text{ext}}(\vec{r}') \right],\\
\delta y(\vec{r}) = -\sum_{\vec{r}'} \left[ G_{yx}(\vec{r} - \vec{r}') \delta f^{x}_{\text{ext}}(\vec{r}') + G_{yy}(\vec{r} - \vec{r}') \delta f^{y}_{\text{ext}}(\vec{r}') \right].
\end{aligned}
\label{Eq_real_space_green}
\end{equation}
\end{small}

The advantage of our technique can be described as follows. There are two constraint equations at each vertex corresponding to forces in the $x$ and $y$ directions. Therefore, in order to solve for the displacements of the $L^2$ vertices in the $x$ and $y$ directions, one needs to simultaneously solve the $2L^2$ constraint equations. A Fourier transform reduces the problem to an inversion of a $2\times2$ matrix at each reciprocal lattice point $\vec{k}$. 

As mentioned above, the elements of the matrix $\text{A}^{-1}$ in Eq.~(\ref{Eq_inverse_matrix}) can be interpreted as the Green's function of the response to a point charge in Fourier space. To obtain simplified expressions for these Green's functions, it is convenient to define the following quantities
\begin{small}
\begin{eqnarray}
\nonumber
\Gamma_1(k_x,k_y,\alpha) &=& -3 - 6\alpha + 2 \alpha \cos(2k_x)\\ 
\nonumber
&&+(3 + 4 \alpha) \cos(k_x) \cos(k_y),\\
\nonumber
\Gamma_2(k_x,k_y,\alpha) &=& -3 - 6\alpha + 
  2(1 + \alpha) \cos(2k_x) \\
\nonumber
&& +(1 + 4\alpha) \cos(k_x) \cos(k_y),\\    
\eta(k_x,k_y,\alpha) &=& \sqrt{3}\sin(k_x)\sin(k_y).
\label{Eq_Green_func0_alpha}
\end{eqnarray}
\end{small}
Here $\alpha$ represents the compression in the system, with lattice constant of the crystalline system given by $R_0 = L_{\text{rest}}(1 +  \alpha)$.
In terms of the above functions, the expressions for the Green's functions simplify to
\begin{eqnarray}
\label{Eq_Green_func_alpha}
\nonumber
 \Tilde{G}_{xx}(k_x,k_y,\alpha) &=& -(1 + \alpha) \frac{\Gamma_1}{(\Gamma_1 \Gamma_2) - {\eta}^2}, \\
\nonumber
\Tilde{G}_{xy}(k_x,k_y,\alpha) &=& -(1 + \alpha) \frac{\eta}{(\Gamma_1 \Gamma_2) - {\eta}^2}, \\ 
\nonumber
\Tilde{G}_{yx}(k_x,k_y,\alpha) &=& -(1 + \alpha) \frac{\eta}{(\Gamma_1 
\Gamma_2) - {\eta}^2}, \\
\Tilde{G}_{yy}(k_x,k_y,\alpha) &=& -(1 + \alpha) \frac{\Gamma_2}{(\Gamma_1 \Gamma_2) - {\eta}^2}. 
\end{eqnarray}

\begin{figure}[t!]
\includegraphics[scale=0.31]{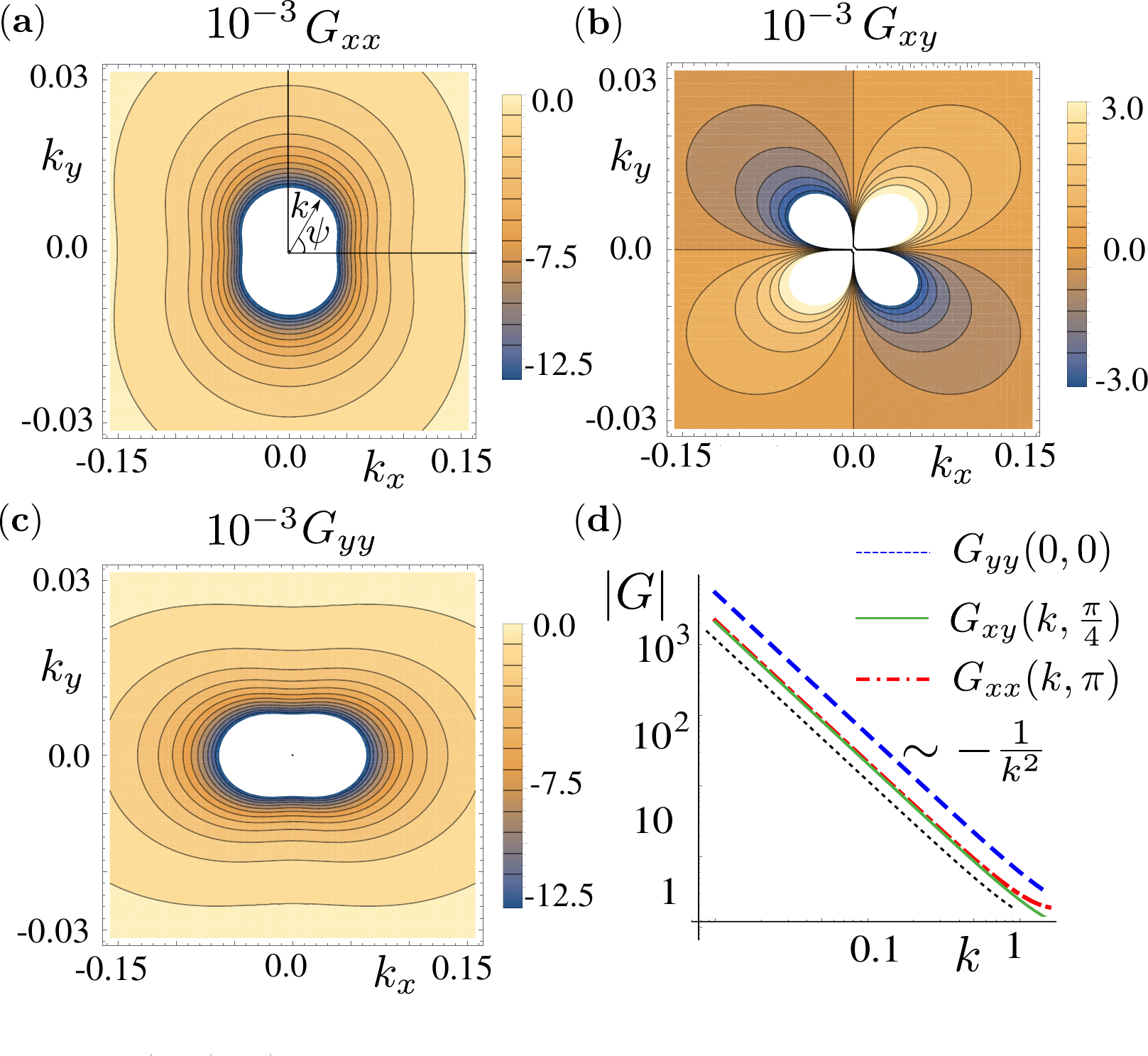}
\caption{The Green's functions of the response $\Tilde{G}_{\mu \nu}(\vec{k})$ in Fourier space $\vec{k} \equiv (k_x, k_y) \equiv (k\cos(\psi),k\sin(\psi))$. {\bf (a)} $\Tilde{G}_{xx}(\vec{k})$ {\bf (b)} $\Tilde{G}_{xy}(\vec{k})${\bf (c)} $\Tilde{G}_{yy}(\vec{k})$. Note that the limits are different along different angles $\psi$ as $k \to 0$.  {\bf (d)} All Green's functions display a $\sim 1/k^2$ behaviour at small $k$.}
\label{fig_greens_func_Fourier}
\end{figure}


\subsection{Continuum Green's Functions}
We next use the framework developed above to derive the continuum behaviour of this system. To obtain the behaviour at large distances $r$, we analyze these expressions at small values of $k$. In the limit $k \to 0$, we obtain the following expression for the Green's functions in Fourier space, with $(k_x,k_y) \equiv (k\cos(\psi),k\sin(\psi))$
\begin{small}
\begin{eqnarray}
   \nonumber
   \Tilde{G}_{xx}(k,\psi) &=& -\frac{1}{k^2}\frac{(2+ 2\alpha)(3 + 8\alpha + 4\alpha \cos(2\psi))}{(3 + 16\alpha + 16 \alpha ^2)(2 + \cos(2\psi))^2}, \\
   \nonumber
   \Tilde{G}_{xy}(k,\psi) &=& -\frac{1}{k^2}\frac{2\sqrt 3(1+\alpha)( \sin(2\psi))}{(3 + 16\alpha + 16 \alpha ^2)(2 + \cos(2\psi))^2},  \\
   \nonumber
 \Tilde{G}_{yy}(k,\psi) &=& -\frac{1}{k^2}\frac{(2 + 2 \alpha)(5+ 8\alpha + 4(1 + \alpha)\cos(2\psi))}{(3 + 16\alpha + 16 \alpha ^2)(2 + \cos(2\psi))^2}.\\
 \label{Eq_small_k}
\end{eqnarray}
\end{small}
The transverse Green's functions of the response are equal with $\Tilde{G}_{yx}(\vec{k}) = \Tilde{G}_{xy}(\vec{k})$. We note that all the Green's functions display a $\sim 1/k^2$ behaviour at small $k$. Therefore, in the small $k$ limit we have $\Tilde{G}_{\mu \nu}(k,\psi) = \Tilde{g}_{\mu \nu}(\psi)/k^2$, where $\Tilde{g}_{\mu \nu}(\psi)$ encodes the angular dependence of these functions. The above expressions make it clear that the response of the medium has a non-trivial dependence on the compression $\alpha$. In Fig.~\ref{fig_greens_func_Fourier} we plot these Green's functions in Fourier space, for small values of $k$. We note that the expressions in Eq.~(\ref{Eq_small_k}) have different limits as $k \to 0$ along different directions. Such singularities are linked to the tensorial nature of the mechanical equilibrium constraints governing the stress tensor \cite{jishnu}.

\subsection{Green's Functions in Real Space} 
\label{subsection_greens_real}

\begin{figure*}[t!]
\includegraphics[scale=0.21]{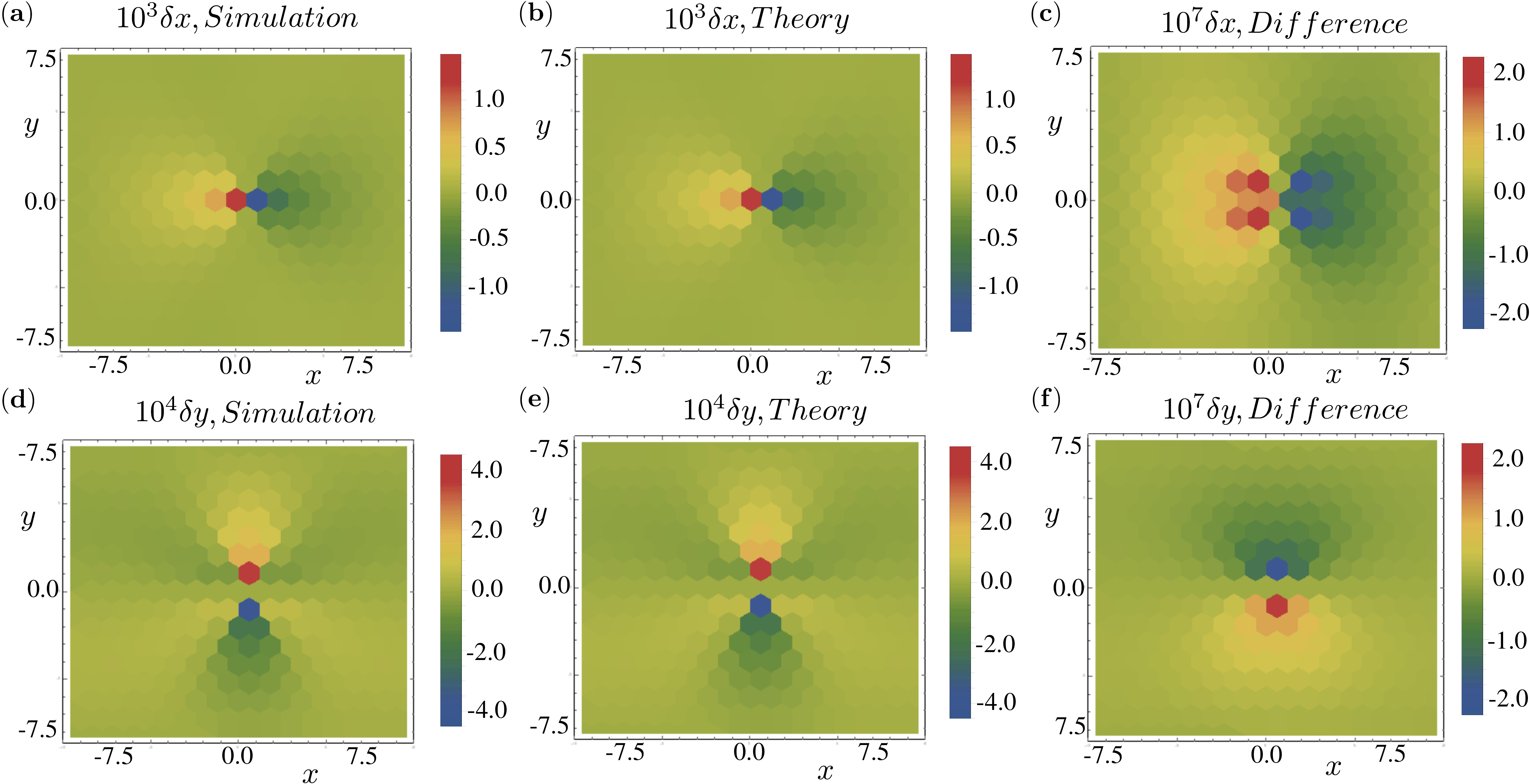}
\caption{Plots of {\bf (a)} the displacements of each particle along the $x$-direction produced by a force dipole obtained from simulations {\bf (b)} $x$-displacements from theory {\bf (c)} difference in $x$-displacements from theory and simulations. {\bf (d)} The displacements along the $y$-direction obtained from simulations {\bf (e)} $y$-displacements from theory {\bf (f)} difference in $y$-displacements from theory and simulations. The external forces are placed on adjacent sites at the origin, with an orientation along the $x$-direction ($\phi = 0$). The magnitude of the force dipole is $f_0$ =0.005.}
\label{Fig_2D_map}
\end{figure*}

We next study the Green's function in real space $\vec{r} \equiv (r,\theta)$, which can be obtained as an inverse Fourier transform of the expressions in Eq.~(\ref{Eq_Green_func_alpha}). In the infinite system size limit $L \to \infty$, Eq.~(\ref{Eq_Green_func_real1}) can be expressed as an integral 
\begin{equation}
\label{Eq_Green_func_real}
    \begin{aligned}
      G_{\mu \nu}(\vec{r}) &=&  \frac{1}{(2\pi)^2} \int_{-\pi}^{\pi}\int_{-\pi}^{\pi} e^{-i\vec{k}\cdot\vec{r}} \Tilde{G}_{\mu \nu}(\vec{k}) d{k_x}d{k_y}.
    \end{aligned}
\end{equation}
We can now use these equations to predict the continuum response at large $r$. We convert this into an integral over the radial and angular coordinates in Fourier space as $(k_x,k_y) \equiv (k\cos(\psi),k\sin(\psi))$.
Using Eq.~(\ref{Eq_small_k}) we have 
\begin{eqnarray}
G_{\mu \nu}(\vec{r}) = \frac{1}{(2\pi)^2} \int_{-\pi}^{\pi} \int_{-\pi}^{\pi} \frac{\Tilde{g}_{\mu \nu}(\psi)}{k^2} \exp(-i \vec{k}. \vec{r}) d^{2}\vec{k},
\end{eqnarray}
which can be simplified to yield
\begin{small}
\begin{eqnarray}
\nonumber
G_{\mu \nu}(\vec{r}) &=& \frac{1}{(2\pi)^2} \int_{0}^{\pi} \int_{-\pi}^{\pi} \frac{\Tilde{g}_{\mu \nu}(\psi)}{k} \exp(-i k r \cos(\theta - \psi)) dk d\psi.\\
\label{eq_green_k}
\end{eqnarray}
\end{small}
Since for point forces $\tilde{f}^x_{\text{ext}}(\vec{k})$ and $\tilde{f}^y_{\text{ext}}(\vec{k})$ are constant fields in Fourier space, the individual Green's functions can be inverted in Fourier space, and represent the solution to the point charge. As the integral over the radial coordinate in Eq.~(\ref{eq_green_k}) diverges as $k \to 0$, we regularize it by adding a constant in the numerator that cancels this divergence. The large distance behaviour can then be derived using the relation
\begin{eqnarray}
\nonumber
\int_0^{\pi} \frac{1 - \exp(i k x)}{k} dk &=& \gamma + \log[\pi x] - \text{CosIntegral}(\pi x) \\
&&~~~~~~~~~  -i \text{SinIntegral}(\pi x),
 \label{Eq_log_integral}
\end{eqnarray}
where $\gamma = 0.5772...$ is the Euler-Mascheroni constant. Using the fact that $\log(r) \gg \text{CosIntegral}(r)$ for large $r$, we have  $G_{\mu \nu} \sim \log(r)$ at large distances $r$. Therefore the predicted displacement fields due to a single unbalanced force in the system diverges at large distances, as force balance is not satisfied. One therefore needs to consider a {\it pair} of Green's functions, i.e. a force dipole, that produces a convergent answer. 


\section{Response to a force dipole}
\label{sec_Results}

Having developed an exact framework for the response of the athermal network to external forces, we apply our theory to the case of a single force dipole. This represents the simplest possibility of externally imposed or active forces that obey the global force balance constraint. 
We model the external dipole as forces $\vec{f}_{p,\text{ext}}$ and $\vec{f}_{q,\text{ext}}$ exerted on two vertices $p$ and $q$ of the lattice. The forces act along an angle $\phi$ with respect to the $x$-direction as shown in Fig.~\ref{fig_dipole_response} {\bf (a)}. To ensure mechanical equilibrium, we have $\vec{f}_{p,\text{ext}} = -\vec{f}_{q,\text{ext}}$. The strength of the dipole is then $|\vec{f}_{p,\text{ext}}| = |\vec{f}_{q,\text{ext}}| = f_0$. The field of external forces is given by
\begin{equation}
\label{Eq_ext_force}
\vec{f}_{i,\text{ext}} = \vec{f_0} (\delta_{ip} - \delta_{iq}).
\end{equation}

We consider the general case of a dipole of length $2d$ centered at the origin, with forces along an angle $\phi = 0$ with respect to the $x$-axis. The two external forces are placed at $\vec{r}_p= - \vec{d}$ and $\vec{r}_q= \vec{d}$ respectively. We can then use Eq.~(\ref{Eq_real_space_green}) to obtain the displacement fields $\delta x(y)(r,\theta)$ at a general position $\vec{r} \equiv (r,\theta)$. For the simple case of $\phi = 0$ illustrated in Figs.~\ref{Fig_2D_map} {\bf(a)}-{\bf(f)}, we have
\begin{eqnarray}
\nonumber
\delta x(r ,\theta) &=& f_0 \left[ G_{xx}(\vec{r} + \vec{d}) - G_{xx}(\vec{r} - \vec{d}) \right],\\
\delta y(r ,\theta) &=& f_0 \left[ G_{yx}(\vec{r} + \vec{d}) - G_{yx}(\vec{r} - \vec{d}) \right].
\label{Eq_dipole_displacement}
\end{eqnarray}
The displacement fields $\delta x(\vec{r})$ and $\delta y(\vec{r})$ generated as a response to the external force dipole can be represented in polar coordinates $\vec{r} \equiv (r,\theta)$, with $r = \sqrt{{x}^2 + {y}^2}$ and $\theta = \tan^{-1} {(\frac{y}{x})}$ respectively as $\delta x(r,\theta)$ and $\delta y(r,\theta)$. In order to characterize the behaviour of these displacement fields we define radially averaged and angular averaged displacement fields as
\begin{eqnarray}
   \nonumber
   {\mathcal{D}_{r}^{x(y)}}(\theta)  = \int_0^{r_{m}} {\delta {x(y)}} (r,\theta)  r dr,\\
   {\mathcal{D}_{\theta}^{x(y)}}(r)  = \int_0^{2\pi} |{\delta {x(y)}} (r,\theta)| d\theta,
   \label{Eq_define_f}
 \end{eqnarray}
\begin{figure}[t!]
\includegraphics[scale=0.38]{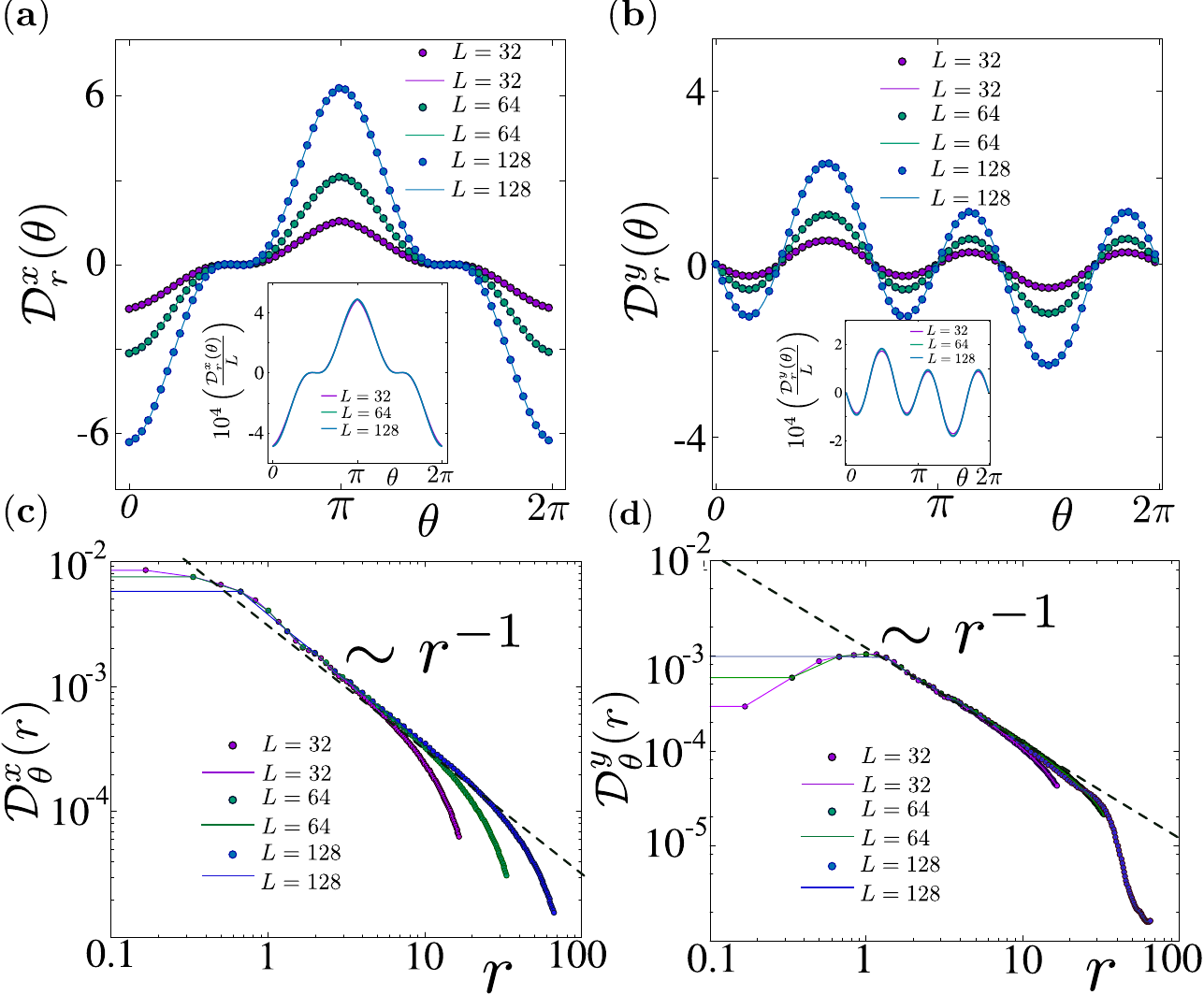}
\caption{Comparison of the radially averaged and angular averaged displacement fields obtained from the theory (solid lines) and numerical simulations (points). 
{\bf (a)} Variation of ${\mathcal{D}_{r}^{x}}(\theta)$ with $\theta$. (Inset) Collapse obtained by scaling with the system size. {\bf (b)} Variation of ${\mathcal{D}_{r}^{y}}(\theta)$ with $\theta$, (Inset) collapse for different system sizes. {\bf (c)} Variation of ${\mathcal{D}_{\theta}^{x}}(r)$ and {\bf (d)} ${\mathcal{D}_{\theta}^{y}}(r)$ with $r$ for different system sizes, displaying a $r^{-1}$ decay at large distances.
The parameters chosen are $L_{\text{rest}} = 1.1$, length of the dipole $2d = 1$ and orientation $\phi = 0$. The strength of the dipole is $f_0 = 0.005$.}
\label{Fig_compare}
\end{figure}
where $r_m = \text{min}[\frac{L_x}{2},\frac{L_y}{2}]$ is the maximum value of $r$ in the system. We calculate these quantities in our  simulations and compare them with results from our theory.
In Fig.~\ref{Fig_compare}, we plot ${\mathcal{D}_{\theta}^{x(y)}}(r)$ and ${\mathcal{D}_{r}^{x(y)}}(\theta)$ for three different system sizes with $L = 32, 64$ and $128$ obtained from numerical simulations. The dipole is placed at locations $(-d,0)$ and $(d,0)$ with $d = 0.5$ . In these simulations, the rest length of the springs is set to $L_{\text{rest}} = 1.1$, while the inter-particle distance in the initial configuration is $R_0 = 1.2$. We also plot results obtained from our theory, which match the numerical simulations exactly. We have also verified our theory for different orientations $\phi$ and lengths $d$ of the dipole as well as for different {\it precompressions} in the lattice (see Supplemental Material for details ~\cite{supplemental}).

\subsection*{Continuum Response to Force Dipole}

\begin{figure}[t!]
\includegraphics[scale=0.2]{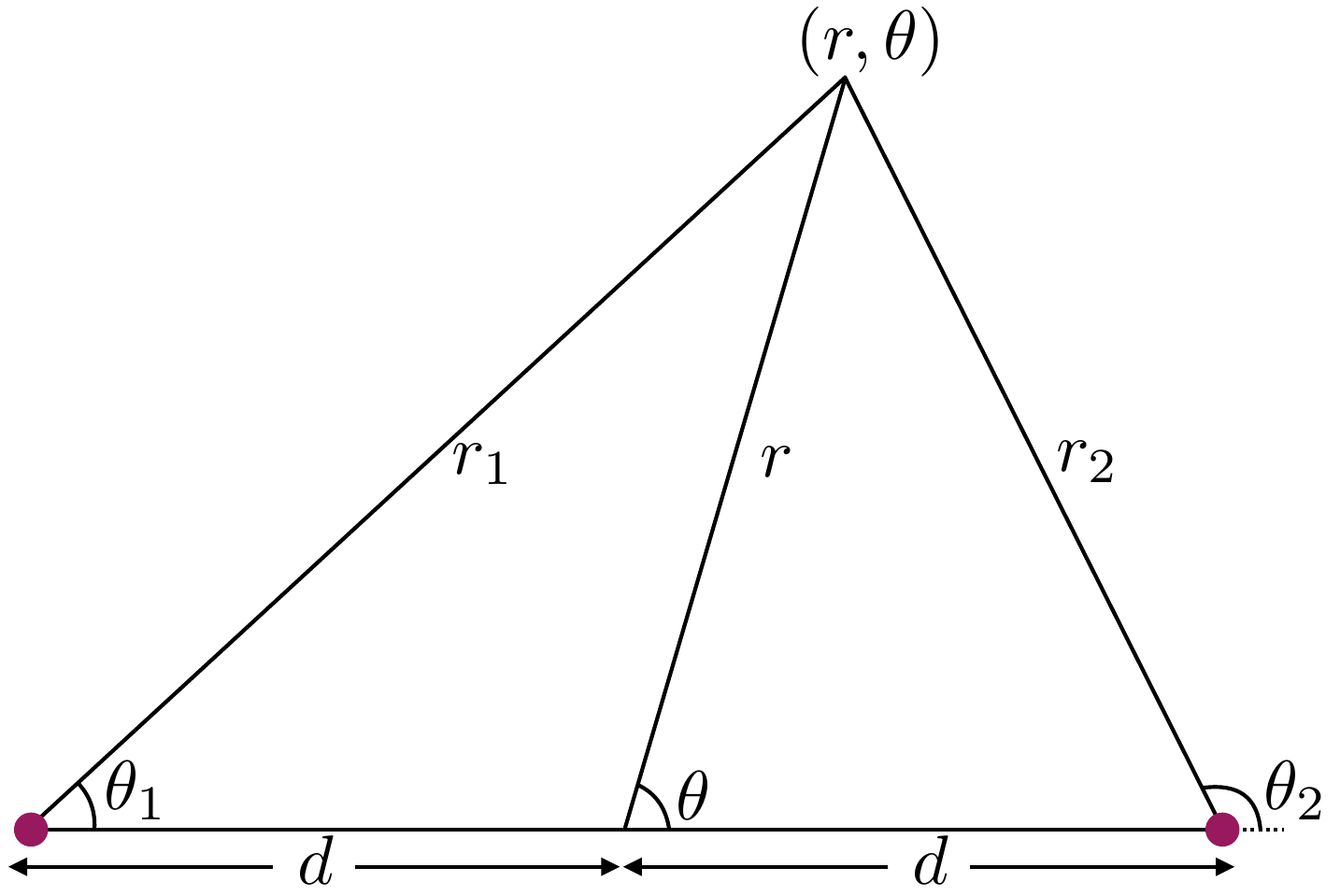}
\caption{Schematic of the geometry used in the computation of displacement fields as a response to a force dipole. The two circles (purple) represent the positions of the external forces which are at a distance $2d$ apart.}
\label{Fig_dipole_notation}
\end{figure}

We next derive the continuum limit of the displacement response to a force dipole. We consider the simplest case of a dipole of length $2d$ centered at the origin $(0,0)$. The field of external forces is then expressed as
\begin{eqnarray}
\nonumber
\delta f^{x}_{\text{ext}}(\vec{r}) &=& f_0  \cos (\phi) [ \delta(\vec{r} - \vec{d}) - \delta(\vec{r} + \vec{d})],\\
\delta f^{y}_{\text{ext}}(\vec{r}) &=& f_0 \sin (\phi) [ \delta(\vec{r}  - \vec{d}) - \delta(\vec{r} + \vec{d})].
\end{eqnarray}
For simplicity, we consider the external forces to be acting along the $x$-direction ($\phi = 0$), however the generalization to non-zero $\phi$ is straightforward. The displacement fields at a distance $\vec{r} = (r,\theta)$ from the center of the dipole (see Fig.~\ref{Fig_dipole_notation}) is then given by
\begin{eqnarray}
\nonumber
\delta x(r ,\theta) &=& f_0  \left[ G_{xx}(\vec{r}_1) - G_{xx}(\vec{r}_2) \right], \\
\delta y(r ,\theta) &=& f_0  \left[ G_{yx}(\vec{r}_1) - G_{yx}(\vec{r}_2) \right]. 
\end{eqnarray}
Next, using the identity in Eq.~(\ref{Eq_log_integral}), and keeping only the real terms, the displacement fields can be expressed as
\begin{eqnarray}
\label{Eq_Large_r_sol}
\nonumber
\delta x(r ,\theta) &=& \frac{2 f_0}{4 {\pi}^2}  \int_{-\pi}^{\pi} \tilde{g}_{xx} 
(\psi) \Big[\log \Big(\frac{r_2 \cos(\theta_2 - \psi)}{r_1 \cos(\theta_1 - \psi)} \Big) \\
\nonumber
&&+\text{CosIntegral}(\pi r_1 \cos(\theta_1 - \psi )) \\
\nonumber
&&- \text{CosIntegral}(\pi r_2 \cos(\theta_2 - \psi)) \Big] d\psi, \\
\nonumber
\nonumber
\delta y(r ,\theta) &=& \frac{2 f_0}{4 {\pi}^2}  \int_{-\pi}^{\pi} \tilde{g}_{yx} 
(\psi) \Big[\log \Big(\frac{r_2 \cos(\theta_2 - \psi)}{r_1 \cos(\theta_1 - \psi)} \Big) \\
\nonumber
&&+\text{CosIntegral}(\pi r_1 \cos(\theta_1 - \psi )) \\
&&- \text{CosIntegral}(\pi r_2 \cos(\theta_2 - \psi)) \Big] d\psi.
\end{eqnarray}

\begin{figure}[t!]
\includegraphics[scale=0.4]{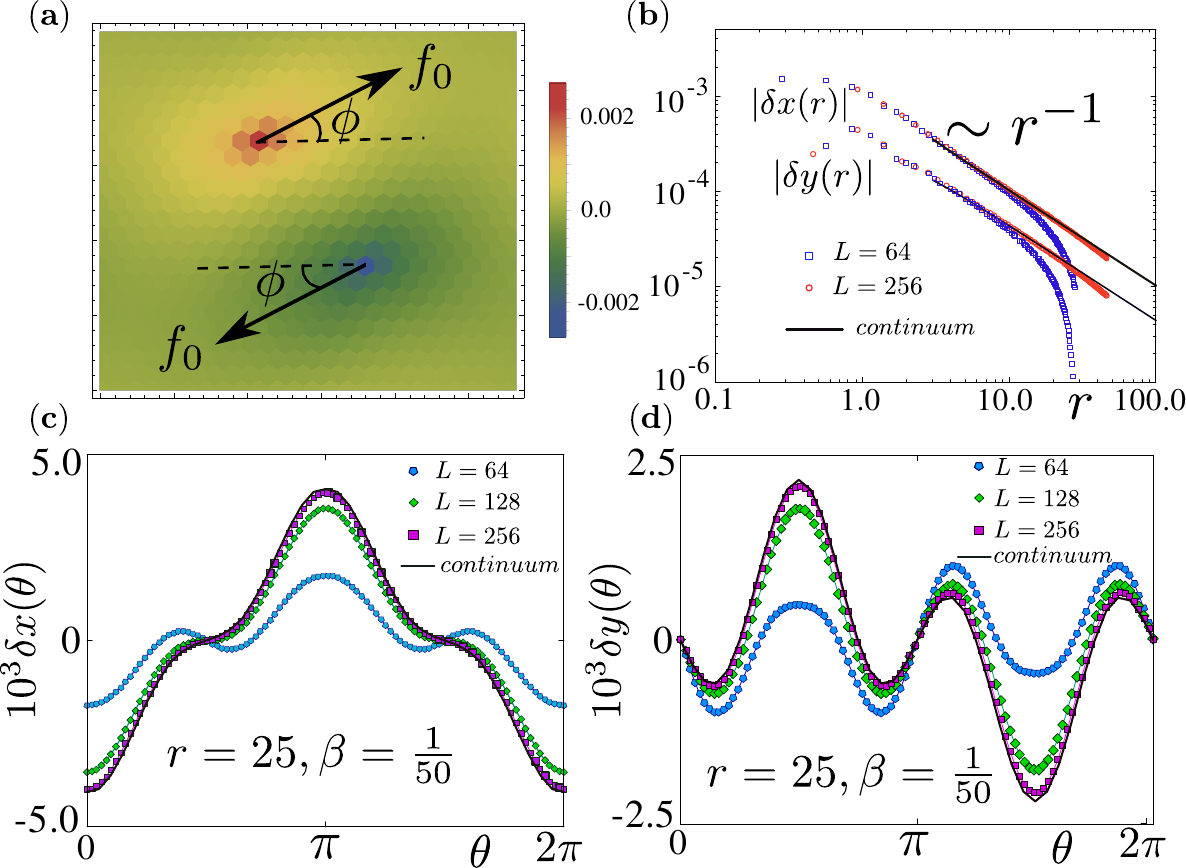}
\caption{{\bf (a)} The displacement $\delta x$ of every vertex along the $x$-axis as a response to external forces at two sites (force dipole) with an orientation $\phi$. {\bf (b)} These displacement fields decay as $1/r$ at large distances away from the dipole.
{\bf (c)} The angular behaviour of the displacement field $\delta x(\theta)$ with $\phi = 0$, and {\bf (d)} $\delta y(\theta)$ at a fixed $r=25$ and $\beta = d/r = 1/50$ for different system sizes $L =64, 128, 256$ obtained from simulations. The predictions from the theory (solid lines) match the simulations exactly. The numerical results converge to the continuum limit predictions (black line) at larger system sizes. Here $\alpha = 1/11$.}
\label{fig_dipole_response}
\end{figure}

Next, we express the distances and angles $r_1$, $r_2$, $\theta_1$, $\theta_2$ in terms of $r$, $\theta$, $d$ (see Fig.~\ref{Fig_dipole_notation}) to
get the following 
\begin{small}
\begin{eqnarray}
\label{Eq_r_theta}
\nonumber
r_1 \cos(\theta_1 - \psi ) &=& r \Big((\cos\theta  + \beta)\cos\psi + \sin\theta \sin\psi \Big), \\
r_2 \cos(\theta_2 - \psi ) &=& r \Big((\cos\theta - \beta) \cos\psi + \sin\theta \sin\psi \Big),
\end{eqnarray}
\end{small}

where $\beta = \frac{d}{r}$. Using Eq.~(\ref{Eq_r_theta}), and using the fact that at large $r$, $\log(r) \gg \text{CosIntegral}(r)$, the limiting solution to the displacement field at large $r$ is given by
\begin{eqnarray}
\label{Eq_final_r_sol}
\nonumber
\delta x(r ,\theta) &=& \frac{2 f_0}{4 {\pi}^2}  \int_{-\pi}^{\pi} d\psi \tilde{g}_{xx} (\psi)
  \Big[\log \left(\frac{\cos(\theta - \psi) - \beta}{\cos(\theta - \psi) + \beta} \right) \Big],\\
\nonumber
\delta y(r ,\theta) &=& \frac{2 f_0}{4 {\pi}^2}  \int_{-\pi}^{\pi} d\psi \tilde{g}_{yx}(\psi) 
 \Big[\log \left(\frac{\cos(\theta - \psi) - \beta}{\cos(\theta - \psi) + \beta} \right) \Big].\\
\end{eqnarray}
As is clear from the above expressions, the $r$ dependence in the displacement fields in the continuum limit arises due to $\beta = d/r$ which emerges as the only relevant lengthscale in the system. Therefore, using the expression for the continuum Green's function in Eq.~(\ref{Eq_final_r_sol}), it is easy to show $\delta x(r) \sim \frac{d}{r}$ and $\delta y(r) \sim \frac{d}{r}$ at large distances $r$ away from the dipole.  In Fig.~\ref{fig_dipole_response} we display the displacement fields obtained using the above theory, which is identical to the response found from simulations. We also display the convergence of the numerical results to the continuum theory predictions as larger system sizes are approached. The changes in inter-particle forces as a response to the external force dipole can now be computed at every bond using Eq.~(\ref{Eq_dipole_displacement}) and the linearized expressions in Eq.~(\ref{Eq_force_linear}).

\section{Randomly Pinned Networks}
\label{sec_pinned_network}

Finally, we turn our attention to randomly pinned athermal networks. Such a situation naturally arises in systems of active particles at high densities where large jammed regions can arise, such as in systems displaying motility induced phase separation \cite{cates2015motility}. In such systems a large collection of particles become ``actively jammed'' with the directions and magnitude of the individual active forces of each particle being randomly distributed, with fixed orientations \cite{rituparno, merrigan2020arrested}. For near-rigid particles, the limit that we are interested in, it is reasonable to assume that the relaxation timescale of the system to settle into a force balanced configuration is much smaller than the timescale of the fluctuations in the directions and magnitude of the forces. Jamming occurs as a result of local as well as global force balance on the network. To model such a situation, we start with a compressed lattice ($\alpha \ne 0$), and external forces $\delta f^{x(y)}_{\text{ext}}(\vec{r})$ at each vertex are chosen from a delta correlated Gaussian distribution such that 
\begin{equation}
\label{Eq_gauss_correlation}
\langle {\delta{f}_{\text{ext}}^{\mu}}(\vec{r})  {\delta{f}_{\text{ext}}^{\nu}}(\vec{r}') \rangle  = \sigma^2   \delta _{\mu \nu} \delta \left(\vec{r} - \vec{r}' \right),
\end{equation}
where the angular brackets $\langle \rangle$ denote the average over realizations of the disorder.
These random forces may lead to a non-zero total force on the system, therefore to ensure global force balance, we impose an additional force  $-1/L^2 \sum_{i = 1}^{L^2} \vec{f}_{\text{ext},i}$ at each vertex. This ensures that $\delta {\tilde{f}_{\text{ext}}^{\mu}(\vec{k}= 0)} = 0$ in Fourier space. The force correlations are then given by
\begin{small}
\begin{eqnarray}
\label{Eq_new_kcorrelation}
\langle {\delta\tilde{
f}_{\text{ext}}^{\mu}}(\vec{k})  {\delta \tilde{f}_{\text{ext}}^{\nu}}(\vec{k}') \rangle  = \sigma^2 \delta_{\mu \nu} \left( \delta \left(\vec{k}  + \vec{k}' \right) - \frac{\delta(\vec{k})  \delta(\vec{k}')}{L^2}  \right).
\label{eq_Fourier_force_correlations}
\end{eqnarray}
\end{small}
The translation invariance of the system ensures that the correlations are non-zero only when $\vec{k} + \vec{k}' = 0$.
Using these external force correlations, we can compute the correlations in the displacement fields
\begin{equation}
\mathcal{C}_{x(y)x(y)}(\vec{r} - \vec{r}') = \langle \delta x(y)(\vec{r}) \delta x(y)(\vec{r}') \rangle.    
\end{equation}
In Fourier space the correlations are given by $\tilde{\mathcal{C}}_{\mu \nu}(\vec{k}) =  \sum_{\vec{r}}  \mathcal{C}_{\mu \nu}(\vec{r})  \exp( {i \vec{k}.\vec{r}})$, therefore $\tilde{\mathcal{C}}_{xx}(\vec{k}) =
\nonumber
\langle \delta \Tilde {x}(\vec{k}) \delta \Tilde {x}(-\vec{k})
\rangle$. Using the expressions in Eq.~(\ref{Eq_Fourier_solution}) and Eq.~(\ref{eq_Fourier_force_correlations}) we have
\begin{eqnarray}
\label{Eq_kcorrelation1}
\tilde{\mathcal{C}}_{\mu\nu}(\vec{k}) =  \sigma^2 \sum_{\alpha}\left[\Tilde{G}_{\mu \alpha}(\vec{k})\Tilde{G}_{\nu \alpha}(-\vec{k}) \right].
\end{eqnarray}
where $\alpha \equiv x,y$. Finally, using the expressions in Eq.~(\ref{Eq_small_k}) with $\Tilde{G}_{\mu \nu}(\vec{k}) \sim \frac{1}{k^2}$ in the limit $k \to 0$, we arrive at the small $k$ behaviour of the correlation functions $\tilde{\mathcal{C}}_{\mu \nu}(\vec{k})  \sim \frac{1}{k^4}$. The correlations in real space can now be computed as an inverse Fourier transform.
Using Eq.~(\ref{Eq_kcorrelation1})  we arrive at the following form for the displacement correlations
\begin{small}
\begin{eqnarray}
\nonumber
\mathcal{C}_{\mu \nu}(\vec{r} - \vec{r}') &=& \frac{\sigma^{2}}{V} \sum_{\vec{k}}\left[\sum_{\alpha}\Tilde{G}_{\mu\alpha}(\vec{k})\Tilde{G}_{\nu\alpha}(-\vec{k}) \right]\exp( -{i \vec{k}.(\vec{r} - \vec{r}')} ),\\
\label{Eq_real_space_correlation}
\end{eqnarray}
\end{small}

\begin{figure}[t!]
\includegraphics[scale=0.21]{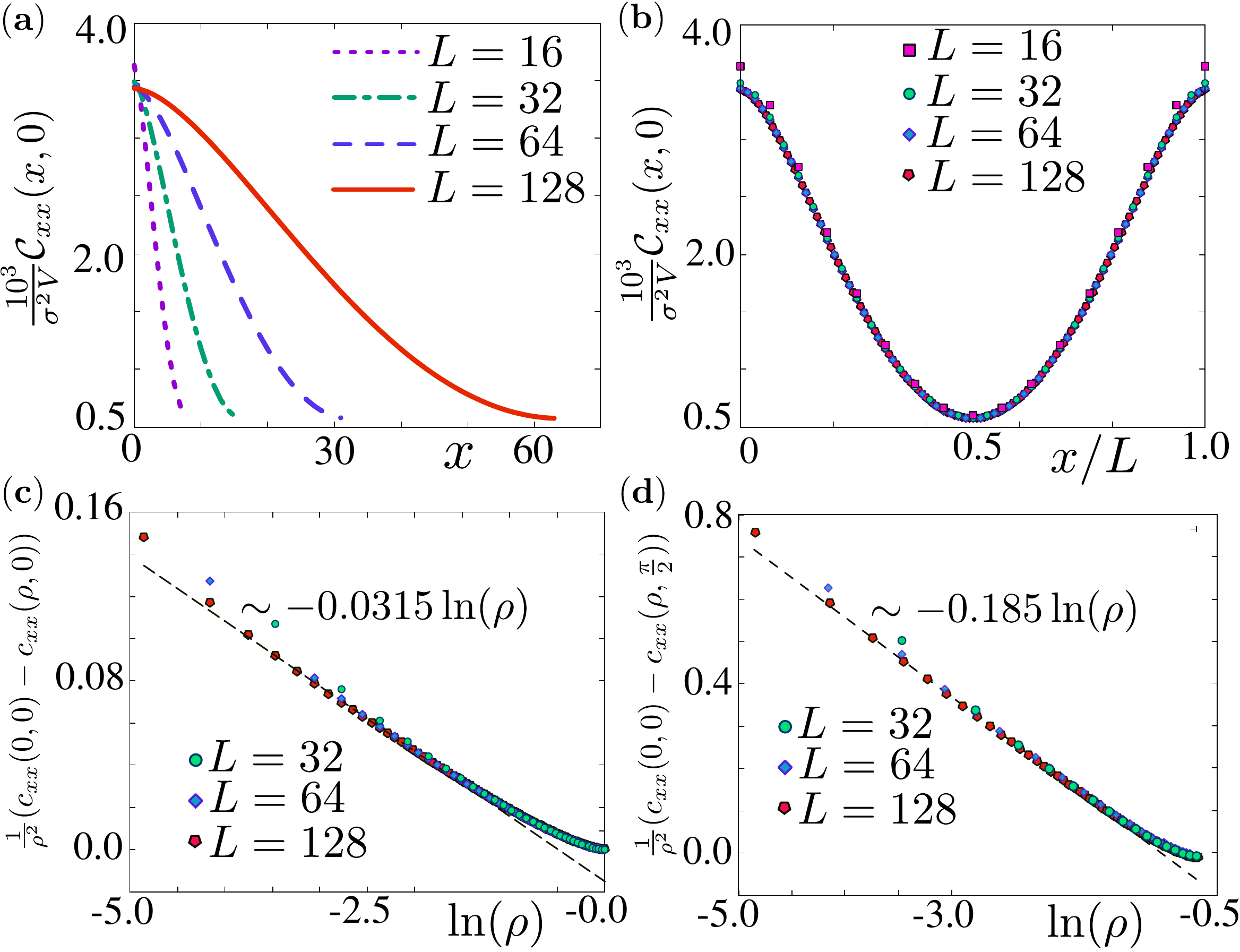}
\caption{Correlations in the displacement fields produced by uncorrelated pinning forces at each site. \textbf{(a)} The $\delta x$ correlations along the $x$ direction  $\mathcal{C}_{xx}(x,0) = \langle \delta x(\vec{r}) \delta x(\vec{r}+x\hat{x}) \rangle$ for different system sizes display long-ranged behaviour,  following the scaling prediction in Eq.~(\ref{eq_scaling_collapse}) as displayed in \textbf{(b)}. The scaled correlation functions $\frac{1}{\rho^2} (c_{xx}(0,0) - c_{xx}({\rho,\theta})$ with $\rho = r/L$ along two different angles \textbf{(c)} $\theta = 0$ and \textbf{(d)} $\theta = \frac{\pi}{2}$. These scaled correlations display different logarithmic corrections along different directions.}
\label{Fig_corr}
\end{figure}

\subsection*{Displacement Correlations in the Continuum Limit}

Finally, we derive continuum limit expressions for the displacement correlations. Transforming the sum in Eq.~(\ref{Eq_real_space_correlation}) to an integral in the $L \to \infty$ limit, the displacement correlation function can be expressed as
\begin{eqnarray}
\nonumber
\mathcal{C}_{\mu\nu}(\vec{r} - \vec{r}') &=& \frac{\sigma^{2}}{(2 \pi)^2} \int_{-\pi}^{\pi} \int_{-\pi}^{\pi}\left[\sum_{\alpha}\Tilde{G}_{\mu\alpha}(\vec{k})\Tilde{G}_{\nu\alpha}(-\vec{k}) \right]  \\
&&~~~~~~~~~~~~\times \exp( -{i \vec{k}.(\vec{r} - \vec{r}')} )  d^{2}{\vec{k}}.
\label{Eq_correlation_integral}
\end{eqnarray}

We can then express Eq.~(\ref{Eq_correlation_integral}) in radial coordinates in Fourier space $\vec{k} \equiv (k\cos(\psi),k\sin(\psi))$ as
\begin{eqnarray}
\nonumber
\mathcal{C}_{\mu\nu}(\vec{r}) &=& \frac{\sigma^{2}}{(2 \pi)^2} \int_{-\pi}^{\pi} d \psi \int_{\epsilon}^{\pi} dk  \left[\sum_{\alpha}\Tilde{G}_{\mu\alpha}(\vec{k})\Tilde{G}_{\nu\alpha}(-\vec{k}) \right]\\
&&~~~~~~~~~~~~~~~~\times \exp( -{i \vec{k}.\vec{r}}).
\label{eq_correlation_intermediate1}
\end{eqnarray}
Since the $\vec{k} = 0$ point is excluded, $\epsilon=\frac{2\pi}{L}\xi$ represents the system-size dependent lower limit in radial coordinates in Fourier space. $\xi$ represents an $\mathcal{O}(1)$ tuning parameter that accounts for the transformation to radial coordinates. 
Next, using the relation $\Tilde{G}_{\mu \nu}(\vec{k}) = \frac{\tilde{g}_{\mu \nu}(\psi)}{k^2}$, and $\tilde{g}_{\mu \nu}(\psi) = \tilde{g}_{\mu \nu}(\pi + \psi)$ we can express the correlations in terms of the angular factors $\tilde{g}_{\mu \nu}(\psi)$ as

\begin{eqnarray}
\nonumber
\mathcal{C}_{\mu\nu}(\vec{r}) &=& \frac{\sigma^{2}}{(2 \pi)^2}  \int_{-\pi}^{\pi} d \psi \left[\sum_{\alpha}\Tilde{g}_{\mu\alpha}(\psi)\Tilde{g}_{\nu\alpha}(-\psi) \right] \\
&&\times \underbrace{\int_{\epsilon}^{\pi} d k \frac{\exp( - i \vec{k}.\vec{r})}{k^{3}}}_{\mathcal{I}(\epsilon,r,\theta,\psi)}.
\label{eq_correlation_intermediate2}
\end{eqnarray}
In order to derive a scaling form for the displacement correlations, we analyze the behaviour of the integral
\begin{eqnarray}
\mathcal{I}(\epsilon,r,\theta,\psi) = \int_{\epsilon}^{\pi} d k \frac{\exp( - i \vec{k}.\vec{r})}{k^{3}},
\end{eqnarray}
in the limit $r \gg 1$ and $\epsilon r \ll 1$.
We perform a variable transformation $\kappa = \frac{k}{\epsilon}$ and $\rho = \frac{r}{L}$. In terms of these variables, the integral can be expressed as
\begin{small}
\begin{eqnarray}
\nonumber
\mathcal{I}(\xi,\rho,\theta,\psi) &=& \frac{L^2}{(2\pi)^2 \xi^2} \int_{1}^{\frac{\pi}{\epsilon}} \frac{\exp(-i2\pi \xi \kappa\rho \cos(\theta - \psi))}{\kappa^3} d\kappa.\\
\label{k4form1}
\end{eqnarray}
\end{small}
In the $\epsilon \to 0$ limit we can extend the limit of the integral $\frac{\pi}{\epsilon} \to \infty$, therefore
\begin{small}
\begin{eqnarray}
\nonumber
\mathcal{I}(\xi,\rho,\theta,\psi) &=&\frac{L^2}{(2\pi)^2 \xi^2} \int_{1}^{\infty} \frac{\exp(-i2\pi \xi \kappa\rho \cos(\theta - \psi))}{\kappa^3} d\kappa.\\
\label{k4form}
\end{eqnarray}
\end{small}
Next, using the identity
\begin{eqnarray}
\nonumber
 \int_{1}^{\infty} d k \frac{e^{-i k \mathcal{R}}}{k^3} &=& \frac{1}{2} + \frac{\mathcal{R}^2}{4} \left(\log \mathcal{R}^2 +2 \gamma -3 \right) + \mathcal{O}(\mathcal{R}^3),\\
\label{k4integration}
\end{eqnarray}
for small $\mathcal{R}$, along with $ \mathcal{R} = 2\pi\rho\xi \cos(\theta - \psi)$, Eq.~(\ref{k4form}) can be expressed as
\begin{small}
\begin{eqnarray}
\nonumber
\mathcal{I}(\xi,\rho,\theta,\psi) &=& \frac{L^2}{(2\pi)^2 \xi^2}  \Big[ \frac{1}{2} + \frac{1}{4} (2\pi \rho \xi)^2 \cos^2(\theta - \psi) \\
\nonumber
&&\times \big( \log(\cos^2(\theta - \psi))
+ 2\log(2\pi\xi \rho) + 2\gamma -3 \big) \Big].\\
\label{continuumform1}
\end{eqnarray}
\end{small}
Substituting this form back into Eq.~(\ref{eq_correlation_intermediate2}) leads to the following scaling form for the correlations
\begin{eqnarray}
\mathcal{C}_{\mu \nu}(\vec{r}) =   \sigma^{2} V c_{\mu \nu}\Big(\frac{r}{L}, \theta \Big) \equiv \sigma^{2} V c_{\mu \nu}\Big(\frac{x}{L},\frac{y}{L} \Big).
\label{eq_scaling_form}
\end{eqnarray}
Using Eq.~(\ref{continuumform1}), the scaled correlations can be shown to have the following form 
\begin{eqnarray}
c_{\mu \nu}(\rho, \theta) &\approx& \text{const}_{\mu \nu} -( \text{a}_{\mu \nu}(\theta) +  \text{b}_{\mu \nu}(\theta) \log \rho ) \rho^2,
\label{eq_scaling_collapse}
\end{eqnarray}
where the coefficients $\text{const}_{\mu \nu}$, $\text{a}_{\mu \nu}(\theta)$, and $\text{b}_{\mu \nu}(\theta)$ depend on the angle $\theta$ at which these correlations are measured (see Supplemental Material for details \cite{supplemental}). 
Finally, to elucidate the nature of the logarithmic terms in the scaled correlation functions, in Figs.~\ref{Fig_corr} ({\bf c}) and ({\bf d}) we plot $\sigma^2\rho^2(c_{xx}(0,0) - c_{xx}(\rho,0))$ as a function of $\log(\rho)$ for two different angles $\theta = 0$ and $\theta = \pi/2$. The asymptotic behaviour in the $\rho = \frac{r}{L} \to 0$ limit displays a logarithmic behaviour with different slopes along different directions. The ratio $\text{b}_{xx}(\pi/2)/\text{b}_{xx}(0) \approx 6$ is consistent with the prediction from the continuum limit expressions.

A surprising aspect of these displacement correlations is their long-range nature, which in addition to scaling with the system size, also diverges as the volume of the system. We can interpret these as arising from the long-range nature of the response to localized forces, as demonstrated by the example of the force dipole. The displacement correlations obtained from the above analysis are plotted in Fig.~\ref{Fig_corr}. We also simulate actively pinned networks by averaging over $1000$ force balanced configurations with uncorrelated external forces for each system size with $\sigma =10^{-4}$ (see Fig.~\ref{fig_schematic}). Our numerically obtained correlations match exactly with the theory developed above (see Supplemental Material for details \cite{supplemental}).

\vspace{0.3cm}
 \section{Discussion}
In this paper we have characterised the response of athermal networks to the presence of external or active forces. Using a triangular lattice arrangement, we developed a Green's function formalism that relates the displacement ``fields'' produced as a response to the external ``charges'' imposed by the active forces. 
This enabled us to derive exact results for the displacement fields and correlations as a response to external force perturbations. Our analytic results demonstrate that {\it uncorrelated} active forces generate long-range correlations in such athermal systems.
These results are also relevant in biological networks, where contractile forces are important in many processes such as wound healing and the motion of cytoskeletons mediated by active internal forces \cite{ronceray2016fiber,ronceray2019fiber,schwarz2013physics}. Our analysis can be generalized to incorporate transverse forces \cite{janevs2019statistical,janevs2019statistical1,gov2004membrane}, as well as different periodic backgrounds, in two as well as three dimensions. The techniques introduced in this paper can also be used to study the response of athermal networks in the presence of disorder, such as in disordered crystals \cite{tong2015crystals, acharya,acharya2021disorder}. Finally, it would be interesting to study the effect of thermal fluctuations on the correlations in such systems in order to understand the emergence of lengthscales associated with amorphous disorder \cite{rainone2020pinching,lerner2018characteristic}.

\section*{Acknowledgments} 
We thank \mbox{Surajit} \mbox{Sengupta}, \mbox{Bulbul} \mbox{Chakraborty}, \mbox{Debasish} \mbox{Chaudhuri}, Subhro Bhattacharjee, Pinaki Chaudhuri, Chandan Dasgupta, Mustansir Barma and Jishnu Nampoothiri for useful discussions. This project was funded by intramural funds at TIFR Hyderabad from the Department of Atomic Energy (DAE).

\bibliographystyle{apsrev4-1} 
\bibliography{stress.bib} 

\clearpage

\begin{widetext}

\begin{appendix}
 
\beginsupplement
\section*{\large Supplemental Material for 
``Long-range correlations in pinned athermal networks''}

In this document we provide supplemental figures and details related to the results presented in the main text.

\maketitle


\subsection{Lattice Notation}
\label{supp_section_linear_theory}

We place the sites of the triangular lattice on a $2 L \times L$ rectangular grid, with the lattice sites $\vec{r} \equiv (x,y)$ such that $\text{mod}(x+y,2)=0$ \cite{horiguchi}. This convention is used in the expressions presented in the main text and supplemental material, while the plotted figures represent our results on the actual triangular lattice. Therefore $\vec{k} = (k_x,k_y) \equiv \Big(\frac{2\pi l}{2L},\frac{2 \pi m}{L}\Big)$ are the reciprocal lattice vectors, and the volume of the system is $V = 2 L^2$. In terms of these vectors, we define the following Fourier coefficients
\begin{eqnarray}
\label{supp_Fourier_coeff}
\nonumber
\mathcal{F}_0(\vec{k}) &=& \exp(-2 i k_x),\\
\nonumber
\mathcal{F}_1(\vec{k}) &=& \exp(-i k_x-i k_y),\\
\nonumber
\mathcal{F}_2(\vec{k}) &=& \exp(i k_x-i k_y),\\
\nonumber
\mathcal{F}_3(\vec{k}) &=& \exp(2 i k_x),\\
\nonumber
\mathcal{F}_4(\vec{k}) &=& \exp(i k_x+i k_y),\\
\mathcal{F}_5(\vec{k}) &=& \exp(i k_y-i k_x).  
\end{eqnarray}
\subsection{Effect of Compression, Orientation and Length of Dipole}
\label{supp_subsection_orientation}
\begin{figure}[ht]
\centering
\includegraphics[scale=0.32]{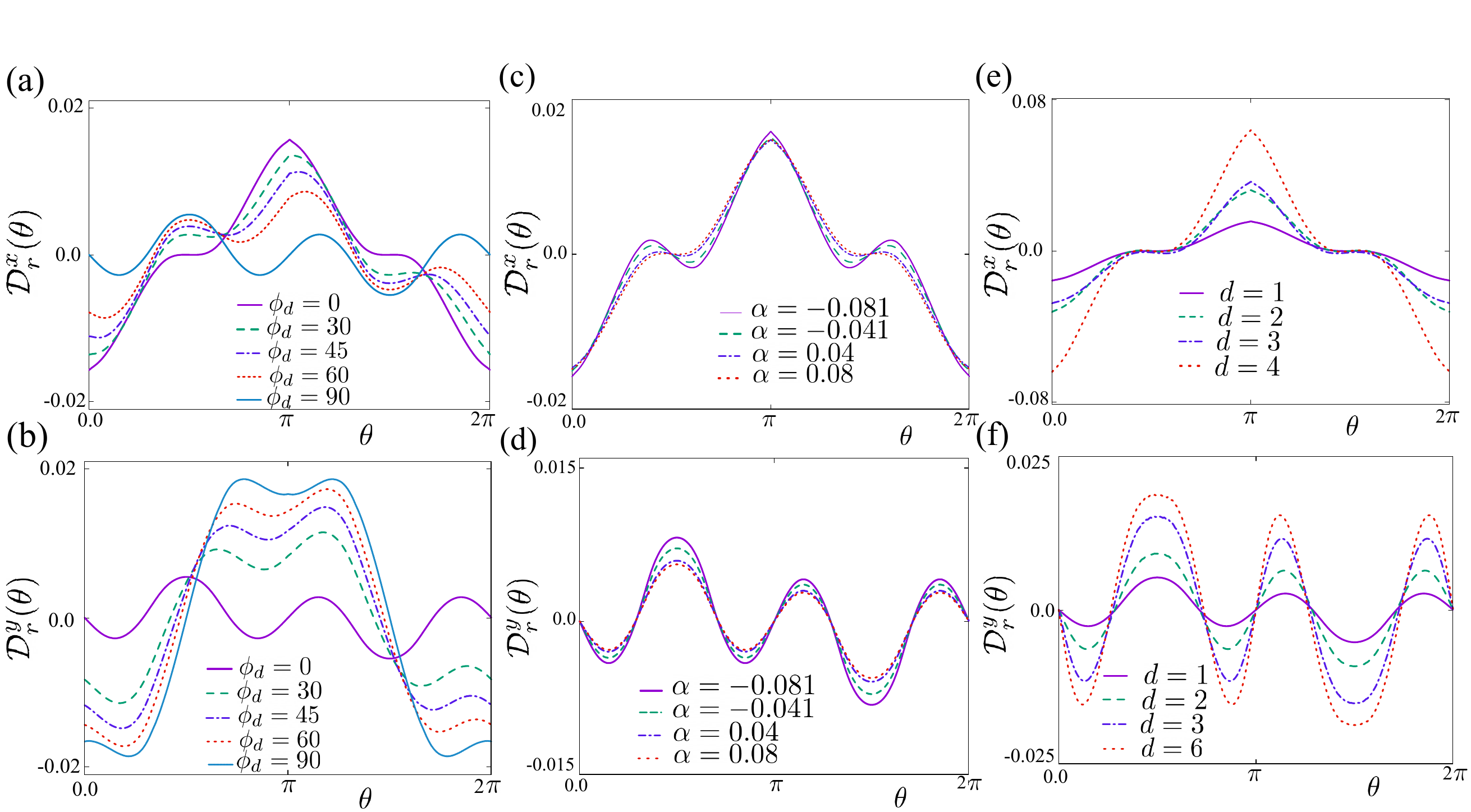}
\caption{{\bf (a)} Plot of the angular averaged displacement fields {\bf (a)} 
${\mathcal{D}_{r}^{x}}(\theta)$ and {\bf (b)}  ${\mathcal{D}_{r}^{y}}(\theta)$ for varying angle $\phi$ of the forces in the dipole, with $\alpha = 1/11$ and $d = 1$. Plot of the angular averaged displacement fields {\bf (c)} ${\mathcal{D}_{r}^{x}}(\theta)$ and {\bf (d)}  ${\mathcal{D}_{r}^{y}}(\theta)$ for different values of the initial compression $\alpha = -0.081, -0.041, 0.04$ and $0.08$ with the angle of the forces $\phi = 0$ and $d = 1$. Plot of the angular averaged displacement fields
{\bf (e)} ${\mathcal{D}_{r}^{x}}(\theta)$ and {\bf (f)}  ${\mathcal{D}_{r}^{y}}(\theta)$ for varying lengths of the dipole $2d$, with $\alpha = 1/11$ and $\phi = 0$.
}
\label{supp_phi}
\end{figure}

We next study the effect of changing the orientation $\phi$ of the forces in the dipole. 
As expected, the $y$-displacements for $\phi = 0^{o}$ are similar to the $x$-displacements for $\phi = 90^{o}$. 

The angular averaged displacement fields are illustrated in Figs.~\ref{supp_rdep} {\bf (a)} and {\bf (b)}. Both ${\mathcal{D}_{\theta}^{x}}(r)$ and ${\mathcal{D}_{\theta}^{y}}(r)$ decay as $r^{-1}$ at large distances $r$ away from the dipole, for all values of $\phi$, consistent with the predictions from our theory. We next study the response of the system by varying the initial compression $\alpha$. In Figs.~\ref{supp_phi} {\bf (c)} and {\bf (d)}, we vary the initial distance $R_0$ between particles, keeping $L_{\text{rest}} = 1.1$ fixed. The angular averaged displacement fields are illustrated in  Figs.~\ref{supp_rdep} {\bf (c)} and {\bf (d)}. Once again, both ${\mathcal{D}_{\theta}^{x}}(r)$ and ${\mathcal{D}_{\theta}^{y}}(r)$ decay as $r^{-1}$ at large distances $r$ away from the dipole, for all values of compression, consistent with the predictions from our theory. Finally, we consider the effect of increasing the length of the dipole $2d$. 
Fig.~\ref{supp_phi} {\bf (e)} and {\bf (f)} display the radially averaged displacement fields as the length of the dipole $d$ is varied.
\begin{figure}[htp]
\centering
\includegraphics[scale=0.3]{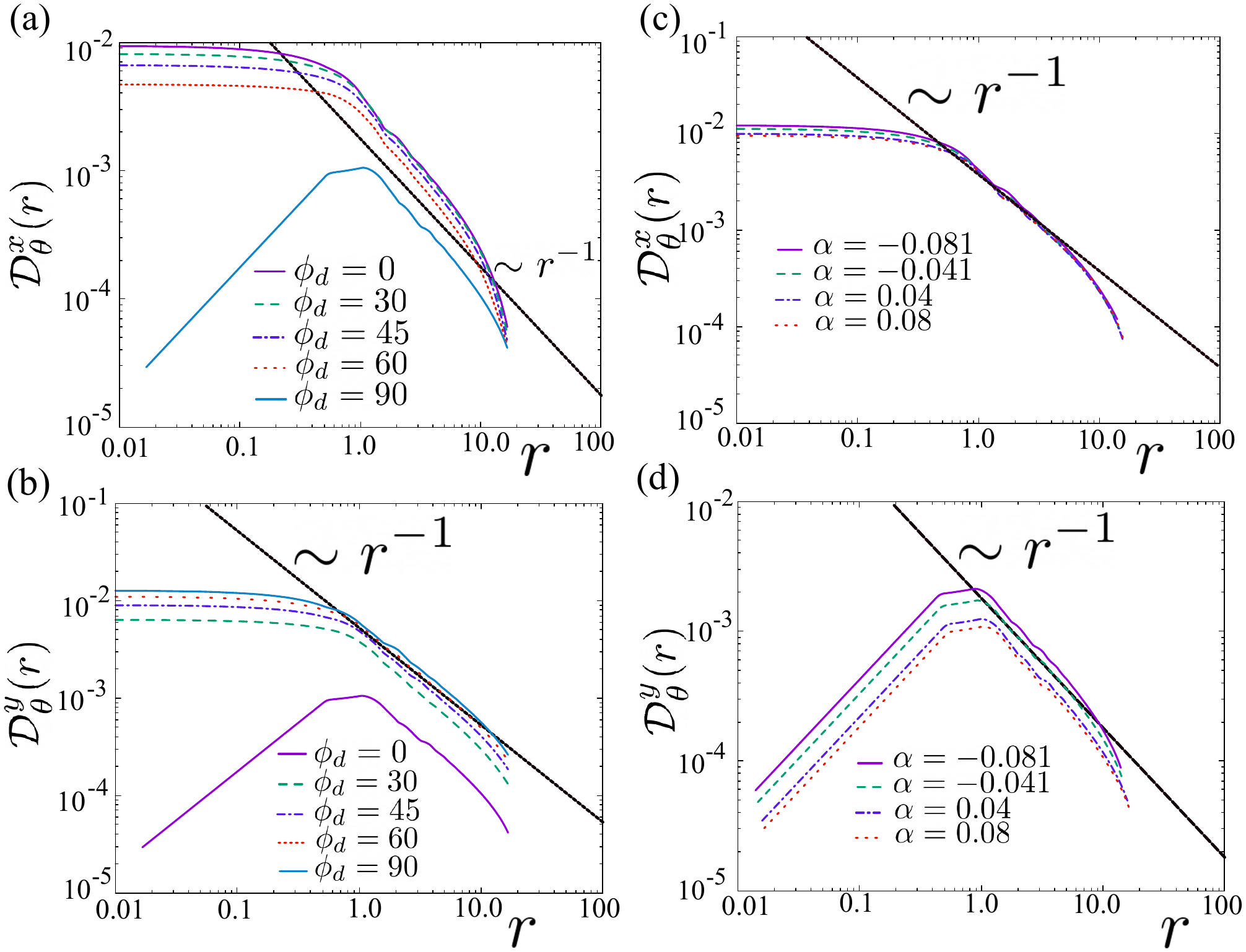}
\caption{The angular averaged displacement fields {\bf (a)}  ${\mathcal{D}_{\theta}^{x}}(r)$ and {\bf (b)} ${\mathcal{D}_{\theta}^{y}}(r)$ for varying angle of the forces in the dipole, with $\alpha = 1/11$ and $d = 1$.
Plot of the angular averaged displacement fields {\bf (c)}  ${\mathcal{D}_{\theta}^{x}}(r)$ and {\bf (d)}  ${\mathcal{D}_{\theta}^{y}}(r)$ for different values of the initial compression $\alpha = -0.081, -0.041, 0.04$ and $0.08$ with the angle of the forces $\phi = 0$ and $d =1$. These fields decay as $r^{-1}$ at large distances $r$ away from the dipole, for all values of the angle $\phi$ and initial compressions $\alpha$, consistent with our theoretical predictions.}
\label{supp_rdep}
\end{figure}
\subsection{Displacement correlations obtained from Simulations}
\label{supp_pinned_athermal}
\begin{figure}[ht!]
\includegraphics[scale=0.23]{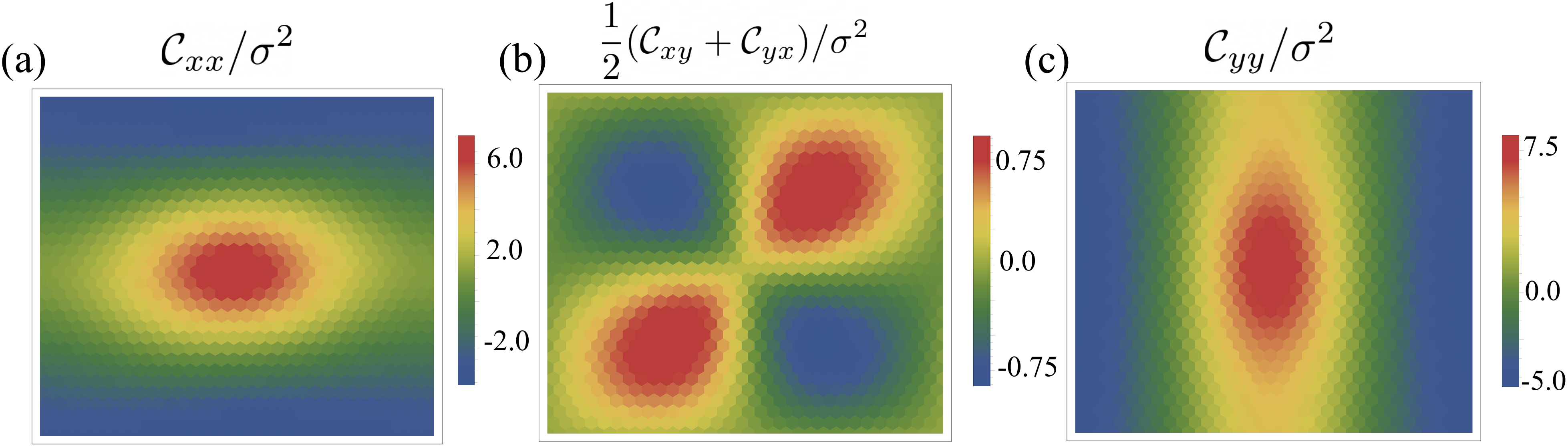}
\caption{Correlations in the displacement fields produced by uncorrelated active forces at each site, obtained from simulations \textbf{(a)} $\mathcal{C}_{xx}(\vec{r}-\vec{r}' ) = \langle \delta x(\vec{r}) \delta x( \vec{r}') \rangle$, \textbf{(b)} $\frac{1}{2}(\mathcal{C}_{xy}(\vec{r}-\vec{r}') + \mathcal{C}_{yx}(\vec{r}-\vec{r}'))= \frac{1}{2} \langle ( \delta x(\vec{r}) \delta y(\vec{r'}) + \delta y(\vec{r}) \delta x(\vec{r'}) \rangle$, \textbf{(c)} $\mathcal{C}_{yy}(\vec{r}-\vec{r}' ) = \langle \delta y(\vec{r}) \delta y( \vec{r'}) \rangle$.
}
\label{supp_correlation}
\end{figure}
The displacement correlations obtained from simulations are displayed in Fig.~\ref{supp_correlation}.
These have been averaged over $1000$ different realizations of energy minimized configurations in the presence of uncorrelated forces, with $\sigma = 10^{-4}$. These match with our theoretical predictions in the main text exactly.

\subsection{Disorder Correlations in Fourier Space}

In the main text, external forces $\delta f^{x(y)}_{\text{ext}}(\vec{r})$ at each vertex $\vec{r}$ are chosen from a delta-correlated Gaussian distribution such that 
\begin{equation}
\label{supp_gauss_correlation}
\langle {\delta{f}_{\text{ext}}^{\mu}}(\vec{r})  {\delta{f}_{\text{ext}}^{\nu}}(\vec{r}') \rangle  = \sigma^2   \delta _{\mu \nu} \delta \left(\vec{r} -\vec{r}' \right).
\end{equation}

In the lattice notation used in the main text, the correlations in the Fourier space can then be expressed as

\begin{eqnarray}
\label{Eq_new_kcorrelation}
\langle {\delta\tilde{
f}_{\text{ext}}^{\mu}}(\vec{k})  {\delta \tilde{f}_{\text{ext}}^{\nu}}(\vec{k}') \rangle  = \frac{\sigma^2}{2} \delta_{\mu \nu} \left( \delta \left(\vec{k}  + \vec{k}' \right)  + \delta\left(\vec{k} + \vec{k}' + (\pi,\pi) \right) \right).
\label{supp_Fourier_force_correlations}
\end{eqnarray}

All functions used in this work display an invariance $\mathcal{F}(k_x,k_y) = \mathcal{F}(\pi - k_x,\pi-k_y)$, arising from the lattice structure. Therefore, for brevity, in the main text we have used
\begin{eqnarray}
\langle {\delta\tilde{
f}_{\text{ext}}^{\mu}}(\vec{k})  {\delta \tilde{f}_{\text{ext}}^{\nu}}(\vec{k}') \rangle  = \sigma^2\delta_{\mu \nu} \delta \left(\vec{k}  + \vec{k}' \right). 
\end{eqnarray}

\subsection{Coefficients in the Continuum Limit}

The scaled displacement correlations in the continuum limit are expressed as
\begin{eqnarray}
c_{\mu \nu}(\rho, \theta) &\approx& \text{const}_{\mu \nu} -( \text{a}_{\mu \nu}(\theta) +  \text{b}_{\mu \nu}(\theta) \log \rho ) \rho^2.
\label{eq_scaling_collapse_supp}
\end{eqnarray}
The coefficients have the following form
\begin{eqnarray}
\nonumber
\text{const}_{xx} & = & \frac{1}{2 \xi^2(2\pi)^4}\int_{-\pi}^{\pi} \big[\Tilde{g}_{xx}(\psi)\Tilde{g}_{xx}(\psi) + \Tilde{g}_{xy}(\psi) \Tilde{g}_{xy}(\psi)\big] d\psi, \\
\nonumber
\text{const}_{xy} & = & \frac{1}{2 \xi^2(2\pi)^4}\int_{-\pi}^{\pi} \big[\Tilde{g}_{xy}(\psi)\Tilde{g}_{yy}(\psi) + \Tilde{g}_{yx}(\psi) \Tilde{g}_{xx}(\psi)\big] d\psi , \\
\text{const}_{yy} &=&  \frac{1}{2 \xi^2(2\pi)^4}\int_{-\pi}^{\pi} \big[\Tilde{g}_{yx}(\psi)\Tilde{g}_{yx}(\psi) + \Tilde{g}_{yy}(\psi) \Tilde{g}_{yy}(\psi)\big] d\psi.
\end{eqnarray}

\begin{small}
\begin{eqnarray}
\nonumber
\text{a}_{xx}(\theta) & = & \frac{1}{2(2\pi)^2} \int_{-\pi}^{\pi} \cos^2(\theta - \psi)\left(\log(|\cos(\theta-\psi)|) + \log(2\pi\xi) + \gamma - \frac{3}{2}\right) \big[\Tilde{g}_{xx}(\psi)\Tilde{g}_{xx}(\psi) + \Tilde{g}_{xy}(\psi) \Tilde{g}_{xy}(\psi)\big] d\psi,\\
\nonumber
\text{a}_{xy}(\theta) & = & \frac{1}{2(2\pi)^2} \int_{-\pi}^{\pi} \cos^2(\theta - \psi)\left(\log(|\cos(\theta-\psi)|) + \log(2\pi\xi) + \gamma - \frac{3}{2}\right) \big[\Tilde{g}_{xy}(\psi)\Tilde{g}_{yy}(\psi) + \Tilde{g}_{yx}(\psi) \Tilde{g}_{xx}(\psi)\big] d\psi,\\
\text{a}_{yy}(\theta) & = & \frac{1}{2(2\pi)^2} \int_{-\pi}^{\pi} \cos^2(\theta - \psi)\left(\log(|\cos(\theta-\psi)|) + \log(2\pi\xi) + \gamma - \frac{3}{2}\right) \big[\Tilde{g}_{yy}(\psi)\Tilde{g}_{yy}(\psi) + \Tilde{g}_{yx}(\psi) \Tilde{g}_{yx}(\psi)\big] d\psi.
\end{eqnarray}
\end{small}
\begin{eqnarray}
\nonumber
\text{b}_{xx}(\theta) & = & \frac{1}{2 (2\pi)^2}\int_{-\pi}^{\pi} \big[\Tilde{g}_{xx}(\psi)\Tilde{g}_{xx}(\psi) + \Tilde{g}_{xy}(\psi) \Tilde{g}_{xy}(\psi)\big]\cos^2(\theta - \psi) d\psi, \\
\nonumber
\text{b}_{xy}(\theta) & = & \frac{1}{2 (2\pi)^2}\int_{-\pi}^{\pi} \big[\Tilde{g}_{xy}(\psi)\Tilde{g}_{yy}(\psi) + \Tilde{g}_{yx}(\psi) \Tilde{g}_{xx}(\psi)\big] \cos^2(\theta - \psi) d\psi , \\
\text{b}_{yy}(\theta) & = & \frac{1}{2 (2\pi)^2}\int_{-\pi}^{\pi} \big[\Tilde{g}_{yx}(\psi)\Tilde{g}_{yx}(\psi) + \Tilde{g}_{yy}(\psi) \Tilde{g}_{yy}(\psi)\big] \cos^2(\theta - \psi) d\psi.
\label{eq_asymptotic_b}
\end{eqnarray}

\end{appendix}
\clearpage

\end{widetext}



\end{document}